\documentclass[sn-mathphys,Numbered]{sn-jnl}

\usepackage{multirow}%
\usepackage{amsmath,amssymb,amsfonts}%
\usepackage{amsthm}%
\usepackage{mathrsfs}%
\usepackage{xcolor}%
\usepackage{textcomp}%
\usepackage{manyfoot}%
\usepackage{booktabs}%
\usepackage{algorithm}%
\usepackage{algorithmicx}%
\usepackage{algpseudocode}%
\usepackage{listings}%
\usepackage[utf8]{inputenc}
\usepackage[english]{babel}
\usepackage[T1]{fontenc}
\usepackage{tikz}
\usepackage{lipsum}

\usepackage{graphicx}
\usepackage{makecell}

\usepackage{amsmath, amssymb, mathtools}
\usepackage{framed} 
\newcommand{\Mod}[1]{\ (\mathrm{mod}\ #1)}

\newcommand{\Supp}{\textsc{Supp}}
\usepackage{physics}
\usepackage{enumitem} 
\usepackage{dashbox}
\usepackage[
    n,
    operators,
    advantage,
    sets,
    adversary,
    landau,
    probability,
    notions,    
    logic,
    ff,
    mm,
    primitives,
    events,
    complexity,
    asymptotics,
    keys]{cryptocode}
    
\newtheorem{theorem}{Theorem}[section]
\newtheorem{corollary}{Corollary}[theorem]
\newtheorem{lemma}[theorem]{Lemma}
\newtheorem{definition}{Definition}[section]

\usepackage{algorithm}
\usepackage{algpseudocode}
\newcounter{protocol}
\makeatletter

\begin{document}

\title{Post-Quantum $\kappa$-to-1 Trapdoor Claw-free Functions from Extrapolated Dihedral Cosets}


\author*[1]{Xingyu Yan}\email{yanxy2020@bupt.edu.cn}

\author*[2]{Licheng Wang}\email{7420220011@bit.edu.cn}

\author[1]{Lize Gu}
\author[3]{Ziyi Li}
\author[1]{Jingwen Suo}
%

%


\affil[1]{State Key Laboratory of Networking and Switching Technology, Beijing University of Posts and Telecommunications, \city{Beijing}, \postcode{100876}, \country{China}}

\affil[2]{School of Cyberspace Science and Technology, Beijing Institute of Technology, \city{Beijing}, \postcode{100081}, \country{China}}


\affil[3]{State Key Laboratory of Information Security, Institute of Information Engineering, UCAS, \city{Beijing}, \postcode{100049}, \country{China}}


\abstract{\emph{Noisy trapdoor claw-free function} (NTCF) as a powerful post-quantum cryptographic tool can efficiently constrain actions of untrusted quantum devices. However, the original NTCF is essentially \emph{2-to-1} one-way function (NTCF$^1_2$). 
In this work, we attempt to further extend the NTCF$^1_2$ to achieve \emph{many-to-one} trapdoor claw-free functions with polynomial bounded preimage size. Specifically, we focus on a significant extrapolation of NTCF$^1_2$ by drawing on extrapolated dihedral cosets, thereby giving a model of NTCF$^1_{\kappa}$ where $\kappa$ is a polynomial integer. Then, we present an efficient construction of NTCF$^1_{\kappa}$ assuming \emph{quantum hardness of the learning with errors (LWE)} problem. We point out that NTCF can be used to bridge the LWE and the dihedral coset problem (DCP). By leveraging NTCF$^1_2$ (resp. NTCF$^1_{\kappa}$), our work reveals a new quantum reduction path from the LWE problem to the DCP (resp. extrapolated DCP). Finally, we demonstrate the NTCF$^1_{\kappa}$ can naturally be reduced to the NTCF$^1_2$, thereby achieving the same application for proving the quantumness. }

\keywords{Noisy trapdoor claw-free functions, Learning with error, Dihedral coset state, Proof of quantumness}



\maketitle

\section{Introduction}\label{sec1}

 The \emph{quantum hardness of LWE} (QLWE) assumes that the LWE problem is hard for any quantum polynomial-time algorithms. Recently, the following connections between the QLWE assumption and quantum computation have received much attention. The first explicit connection is related to the \emph{dihedral coset problem} (DCP). The DCP is quantumly derived from the \emph{dihedral hidden subgroup problem} (DHSP) \cite{ettinger2000quantum,kuper05}. At FOCS '02, Regev \cite{Reg02} proved the first quantum reduction from the lattice problem to the DCP. Namely, if there exists a polynomial-time quantum algorithm that solves the DCP, then there exists an algorithm to solve the $\poly$-unique SVP. Furthermore, Brakerski et al. \cite{BKSW18} at PKC ’18 further generalized Regev's proof and characterized the QLWE assumption under quantum polynomial-time reductions, showing that the LWE problem is quantumly equivalent to a generalization of DCP known as \emph{extrapolated dihedral coset problem} (EDCP).
 
\begin{figure}[t]
\centering
\includegraphics[width=0.95\textwidth]{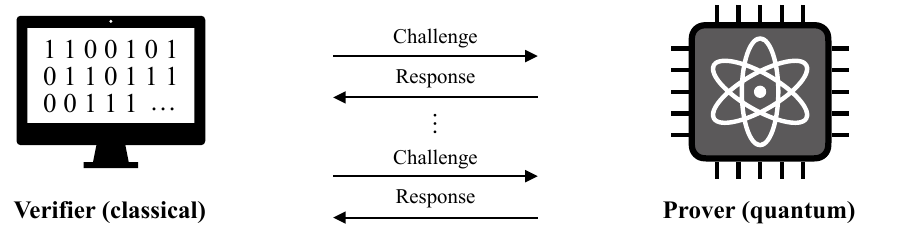}
\caption{Schematic representation of classically verifiable test of quantum devices. Interactions are enabled by using post-quantum cryptographic primitives between a classical verifier and a quantum prover \cite{BCMVV18}.} \label{fig.1}
\end{figure}

 Another interesting connection, initiated by the recent work of Brakerski et al. \cite{BCMVV18} at FOCS '18, is to leverage the QLWE assumption for demonstrating ``proofs of quantumness'' \footnote{The term ``proofs of quantumness'' also known as ``quantum supremacy'', is to demonstrate the quantum computational advantage.}. Therein, an interactive cryptographic protocol was proposed between a polynomial-time \emph{classical verifier} and a single untrusted polynomial-time bounded \emph{quantum prover}, where a quantum prover can successfully respond to the verifier's challenges and will be accepted, whereas any classical prover will be rejected. The interaction model is shown diagrammatically in Fig. \ref{fig.1}. The core subroutine of their work is known as a \emph{qubit certification test protocol}, which extremely relies on a powerful cryptographic tool called the \emph{2-to-1 noisy trapdoor claw-free functions (NTCF$^1_2$)} with \emph{adaptive hardcore bit} (AHB) property.

For now, the 2-to-1 trapdoor claw-free functions (TCF$^1_2$) promise to be an effective tool for limiting, describing, and verifying the actions of untrusted quantum devices. Conceptually, the TCF$^1_2$ is a pair of injective functions $f_0,f_1$ which have the same image, and with access to a secret trapdoor it is easy to evaluate the two preimages $x_0$ and $x_1$ of the same image $y$ such that $f_0(x_0)=f_1(x_1)=y$. However, it is computationally difficult to invert $f_0,f_1$ without the trapdoor. Such a pair of $(x_0,x_1)$ is known as a claw, hence the name is claw-free \cite{Mah18a,Mah18b}. Moreover, the additional AHB property states that whenever $f_0(x_0)=f_1(x_1)$, finding even a single bit of formation about $x_1$ should be computationally intractable for any $x_0$. Note that a TCF$^1_2$ with the AHB property is often referred to as a \emph{strong TCF}, while a TCF$^1_2$ without the AHB property is called as a regular (or ordinary) TCF$^1_2$. Unfortunately, it is unclear how to construct such a \emph{clean} version of strong TCF$^1_2$ under standard cryptographic assumptions. Instead, a \emph{noisy} version of TCF$^1_2$ called NTCF$^1_2$ based on the QLWE assumption is first introduced by \cite{BCMVV18,Mah18b}, which is the only known instance that satisfies the strong TCF$^1_2$ property.

In \cite{BCMVV18}, the qubit certification test essentially employs the NTCF$^1_2$ to enforce qubits structure of the quantum prover. Roughly speaking, the classical verifier uses the NTCF$^1_2$ to enable the quantum prover to create a superposition as $\frac{1}{\sqrt{2}}(\ket{0}\ket{\boldsymbol{x}_0}+\ket{1}\ket{\boldsymbol{x}_0-\boldsymbol{\rm s}}$, where $\boldsymbol{\rm s} \in \ZZ^n_q$ is a fixed secret, $\boldsymbol{x}_0 \in \ZZ^n_q$ is arbitrary and $\boldsymbol{x}_1 = \boldsymbol{x}_0-\boldsymbol{\rm s} \mod q$. Afterward, the verifier randomly chooses to either run a \emph{preimage test} or an \emph{equation test}. In the preimage test, the quantum prover would measure the above state in the computational basis, thereby resulting in 1-bit information of randomness $b\in \bin$, as well as a random preimage, $\boldsymbol{x}_0$ or $\boldsymbol{x}_1$. Instead, in the equation test, the quantum prover would measure the above state in the Hadamard basis, yielding a bit $c$ and a string $d\in \bin^{n\log q}$, which will be checked by the verifier for $c=d\cdot(\boldsymbol{x}_0 \oplus \boldsymbol{x}_1)$ holds \footnote{In fact, the bit $c$ is evaluated by $c=d\cdot(\mathcal{J}({\boldsymbol{x}}_0) \oplus \mathcal{J}({\boldsymbol{x}}_1))$ in \cite{BCMVV18}, where $\mathcal{J}(\cdot)$ is the binary representation function. For simplicity, we omit this function in the expression.}. Notably, to completely protect against a cheating quantum prover, the proposal requires the aforementioned AHB property, which can be informally reformulated as follows: if a quantum polynomial-time bounded prover can pass both the preimage test and the equation test simultaneously, then the quantum prover can break the QLWE assumption by learning a bit of the secret $\boldsymbol{\rm s}$.

Unfortunately, as the NTCF$^1_2$ are essentially 2-to-1 functions, the rate of randomness generation per round is inefficient. To handle this issue, Mahadev et al. \cite{MVV22} recently consider replacing the NTCF$^1_2$ by a powerful cryptographic tool that looks like $k$-to-1 functions for large $k$, with the goal of producing $\log k$ bits of randomness per round as opposed to 1. Concretely, the $k$-to-1 functions introduced in their work arise from the use of leakage resilience techniques of LWE. The main idea behind in \cite{MVV22} is to change the preimage test of the original qubit certification test protocol by swapping the use of an LWE matrix $\boldsymbol{\rm A}$ for a computationally indistinguishable lossy matrix $\boldsymbol{\rm \widetilde{A}}$, where $\boldsymbol{\rm \widetilde{A}} =\boldsymbol{\rm {B}}\boldsymbol{\rm {C}}+\boldsymbol{\rm {E}}$ and $\boldsymbol{\rm {C}}\in \ZZ^{\ell \times n}_q (\ell\approx n / \log q)$. Since the matrix $\boldsymbol{\rm \widetilde{A}}$ is lossy, then for a valid image reported by the quantum prover there are up to $k\approx q^{n-\ell}$ valid preimages. Hence, by exploiting the $k$-to-1 functions based on leakage resilience, their protocol could produce $\bigOmega{n}$ bits of randomness in constant rounds. 

However, in their work, the construction of $k$-to-1 claw-free hash functions based on leakage resilience faces two tricky problems during the equation test phase. Firstly, as the superposition over many preimages instead over two, it is unclear how to perform the Hadamard measurement in the equation test. Secondly, the classical verifier can no longer rely on the knowledge of a trapdoor, since the image $\boldsymbol{y}$ inherently has many valid preimages. In the original qubit certification protocol, we know that the trapdoor is necessary for the verifier to inverse a given image $\boldsymbol{y}$, thereby obtaining the $\boldsymbol{x}_0$ and checking $c=d\cdot(\boldsymbol{x}_0 \oplus (\boldsymbol{x}_0 - \boldsymbol{\rm s}))$ holds. To address both problems simultaneously, Mahadev et al. introduced a hypothetical protocol with a \emph{quantum verifier} who is allowed access to the quantum prover's state to remove the need for the trapdoor. Finally, to revert back to a classical verifier protocol, they proposed a compromise \emph{hybrid argument}, in which the proposal uses the original NTCF$^1_2$ for the test rounds while uses the $k$-to-1 functions for the generation rounds. The key to combining both is that lossy and uniform matrices are computationally indistinguishable. Although their final protocol can be reduced to a classical verifier protocol, the hypothetical quantum verifier introduced during their analysis makes the whole hybrid solution look unintuitive. Based on this, it is worth considering the following natural question:

\begin{center}
    \emph{Assuming the QLWE assumption, can we construct a many-to-one claw-free function and require that this function even have a trapdoor?}
\end{center}

For this purpose, instead of building $k$-to-1 functions via the leakage resilience, we attempt to extend the original NTCF$^1_2$ and directly construct a family of \emph{$\kappa$-to-1 noisy trapdoor claw-free functions}  (NTCF$^1_{\kappa}$) with $\kappa = \poly$ under QLWE assumption. Notably, for constructing NTCF$^1_{\kappa}$, we manifest an interesting connection between the NTCF$^1_{\kappa}$ and the \emph{Extrapolated Dihedral Coset State} introduced in \cite{BKSW18}. Finally, we demonstrate the NTCF$^1_{\kappa}$ can naturally be reduced to the NTCF$^1_2$, thereby achieving the same application for proving the quantumness as \cite{BCMVV18}.

\subsection{Our Contributions}\label{sec1.1}

In this paper, we study a generalization of 2-to-1 NTCF and propose the construction of $\kappa$-to-1 NTCF with $\poly[]$-bounded preimage size from extrapolated dihedral cosets. To summarize, we make the following contributions.

\begin{enumerate}
\item We generalize the NTCF$^1_2$ and extend it to a NTCF$^1_{\kappa}$ family with $\kappa = \poly$, based on the assumption of sub-exponential QLWE of $n$ dimension. Our solution manifests an interesting connection between the NTCF$^1_{\kappa}$ and the extrapolated dihedral coset state. 

\item  We present a new quantum reduction from the LWE to the DCP (resp. EDCP) by using the NTCF$^1_2$ (resp. NTCF$^1_{\kappa}$) shown in Fig. \ref{fig.2}, where NTCF as a bridge plays the key role to efficiently link the LWE and the DCP. 


\begin{figure}[t]
\centering
\includegraphics[width=0.8\textwidth]{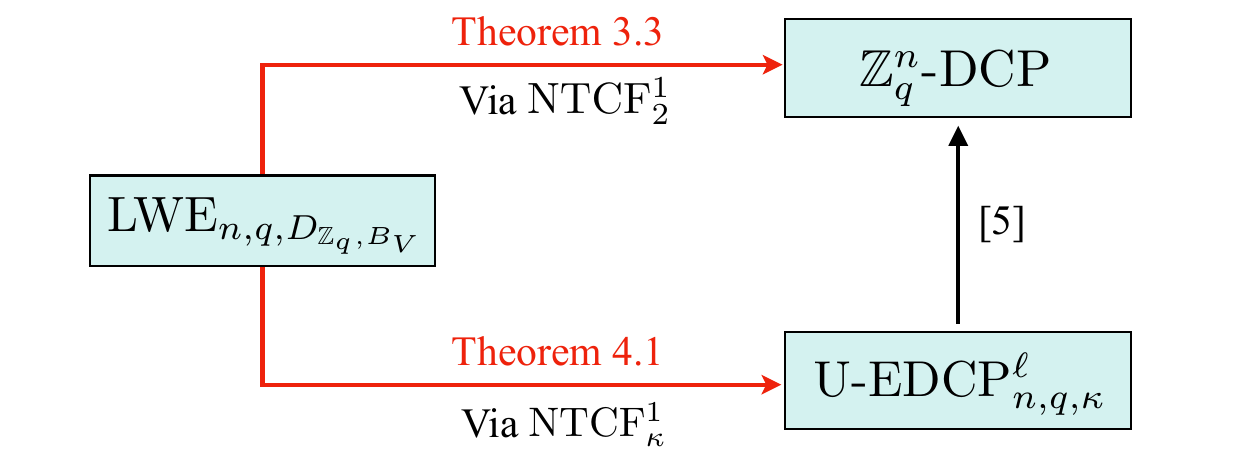}
\caption{Graph of reductions between the LWE problem with super-polynomial modulus and the DCP and the EDCP. The parameters $n$ (the dimension), $q$ (the modulus), $B_V$, $\ell$ and $\kappa$ are functions of security parameter $\lambda$. We assume that $n,\ell$ and $\kappa$ are polynomial in $\lambda$, $q$ is super-polynomial in $\lambda$.} \label{fig.2}
\end{figure}

\end{enumerate}


Informally, the main results are concluded as the following theorems.

\begin{theorem}[informal]\label{theo1}
    There exists a function family $\mathcal{\widetilde{F}}_{\mathrm{LWE}}$ that is a $\kappa$-to-1 noisy trapdoor claw-free functions (NTCF$^1_{\kappa}$) family with $\kappa = \poly$ under the hardness assumption \rm{LWE}$_{n,q,D_{\mathbb{Z}_q,B_L}}$. 
\end{theorem}

\begin{theorem}[informal]\label{theo2}
    Let $\mathcal{F}_{\mathrm{LWE}}$ be a NTCF$^1_2$ family under the QLWE assumption. If there exists a solution to the $\ZZ^n_q$-DCP, then there exists a quantum algorithm for the LWE$_{n,q,D_{\mathbb{Z}_q,B_V}}$ problem with super-polynomial modulus $q$. 
\end{theorem}
 
\begin{corollary}
Let $\mathcal{\widetilde{F}}_{\mathrm{LWE}}$ be a family of NTCF$^1_{\kappa}$ under the QLWE assumption. If there exists an algorithm that solves U-EDCP$^{\ell}_{n,q,\kappa}$, then there exists a solution to the LWE$_{n,q,D_{\mathbb{Z}_q,B_V}}$ problem with super-polynomial modulus $q$. 
\end{corollary}

\begin{theorem}[informal]\label{theo3}
    Let $\mathcal{\widetilde{F}}_{\mathrm{LWE}}$ be a family of NTCF$^1_{\kappa}$ satisfying Theorem \ref{theo1}. There exists a proof of quantumness protocol (without proof of certifiable randomness) based on $\mathcal{\widetilde{F}}_{\mathrm{LWE}}$ consisting of polynomial-time quantum circuits and 2-round communication. 
\end{theorem}

\subsection{Technical Overview}\label{sec1.2}

We begin by reviewing the Brakerski's proof of quantumness protocol \cite{BCMVV18} and the Mahadev’s protocol \cite{MVV22} for certifiable randomness generation. The core ingredient of these protocols is called as the \emph{qubit certification protocol}. Note that the verifier is a \(\mathsf {PPT}\) machine, while the prover is a \(\mathsf {QPT}\) machine.

\subsubsection{Recap: Qubit Certification Protocol and Lossy Randomness Protocol}

  For clarity, we first roughly describe how the QLWE assumption can be employed to construct an NTCF$^1_2$. Consider the domain is $\mathcal{X}$ and the range is $\mathcal{Y}$. Let $\boldsymbol{x}\in \mathbb{Z}^n_q$, and $\boldsymbol{y}\in \mathbb{Z}^m_q$. Informally, given an LWE instance $(\boldsymbol{\rm A},\boldsymbol{\rm t}=\boldsymbol{\rm{As}}+\boldsymbol{\rm e}) \in \mathbb{Z}^{m\times n}_q \times \mathbb{Z}^{m}_q$, it is easy to define an NTCF$^1_2$ family $\mathcal{F}_{\mathrm{LWE}}$ by letting $f_0(\boldsymbol{x})=\boldsymbol{\rm A}\boldsymbol{x}+\boldsymbol{\rm e}_0$ and $f_1(\boldsymbol{x})=\boldsymbol{\rm A}\boldsymbol{x}+\boldsymbol{\rm e}_0+\boldsymbol{\rm t}$. Since $\boldsymbol{\rm t}=\boldsymbol{\rm{As}}+\boldsymbol{\rm e}$, we can see that $f_1(\boldsymbol{x})=\boldsymbol{\rm A}(\boldsymbol{x}+\boldsymbol{\rm s})+\boldsymbol{\rm e}_0+\boldsymbol{\rm e}$. Note that if $\boldsymbol{\rm e}$ is 0, then $f_1(\boldsymbol{x})=f_0(\boldsymbol{x}+\boldsymbol{\rm s})$ holds. Indeed, if $\boldsymbol{\rm e}_0$ sampled from a Gaussian is much wider than $\boldsymbol{\rm e}$, then we can ensure that the distributions $f_1(\boldsymbol{x})$ and $f_0(\boldsymbol{x}+\boldsymbol{\rm s})$ are statistically close, thereby ensuring that $f_1(\boldsymbol{x})=f_0(\boldsymbol{x}+\boldsymbol{\rm s})$. Specifically, according to \cite{BCMVV18}, it is given that $\boldsymbol{\rm e}_0$ and $\boldsymbol{\rm e}$ are drawn independently from distributions $D_{\mathbb{Z}^m_q,B_P}$ and $D_{\mathbb{Z}^m_q,B_V}$ respectively, and $B_P/B_V$ is super-polynomial, so $\|\boldsymbol{\rm e}_0\|\leq B_P\sqrt{m}$, $\|\boldsymbol{\rm e}\|\leq B_V\sqrt{m}$ and $\|\boldsymbol{\rm e}\| \ll \|\boldsymbol{\rm e}_0\|$. Therefore, a \(\mathsf {QPT}\) prover can use an NTCF$^1_2$ to create a superposition as (omitting normalization factors):
\begin{align*}
    \sum_{b\in\{0,1\}}\sum_{\boldsymbol{x}\in \mathbb{Z}^n_q}\sum_{\boldsymbol{\rm e}_0\in \mathbb{Z}^m_q}\ket{b}_{\mathsf B}\ket{\boldsymbol{x}}_{\mathsf X}\ket{\boldsymbol{\rm A}\boldsymbol{x}+\boldsymbol{\rm e}_0+b \cdot \boldsymbol{\rm t}}_{\mathsf Y},
\end{align*}
After measuring the $\mathsf {Y}$ register, the prover returns an image $\boldsymbol{y}$ and the desired claw superposition in $\mathsf{BX}$ registers as
 \begin{align*}
    &\frac{1}{\sqrt{2}}(\ket{0}\ket{\boldsymbol{x}_0}+\ket{1}\ket{(\boldsymbol{x}_0-\boldsymbol{\rm s}) \mod q}),
 \end{align*}
where $\boldsymbol{x}_1=(\boldsymbol{x}_0-\boldsymbol{s}) \mod q$ and $(\boldsymbol{x}_0,\boldsymbol{x}_1)$ form a claw of the $\mathcal{F}_{\mathrm{LWE}}$. Based on this, we now describe the qubit certification protocol, shown in \emph{Protocol 1}. All setting of parameters is the same as the LWE-based NTCF$^1_2$ in Section \ref{sec2.3}. 

\paragraph{\textbf{Protocol 1: Qubit Certification Test}} \label{Proto1}

\begin{enumerate}
  \item The verifier generates $k=(\boldsymbol{\rm A},\boldsymbol{\rm A}\boldsymbol{\rm s}+\boldsymbol{\rm e})\in \mathbb{Z}^{m\times n}_q \times \mathbb{Z}^{m}_q$, along with the trapdoor $t_{\boldsymbol{\rm A}}$.
  
  \item The verifier sends $k$ to the prover.
  
  \item The prover reports an image ${\boldsymbol{y}}\in \mathbb{Z}^{m}_q$ to the verifier.
  
  \item The verifier randomly selects to either run a \emph{generation round} ($C=G$) or a \emph{test round} ($C=T$) and sends $C$ to the prover.
    \begin{enumerate}

    \item \textbf{$C=G$}:
    The verifier runs the preimage test: the prover is asked to return a bit $b$ and a corresponding preimage $\boldsymbol{x}$ of $\boldsymbol{y}$. The verifier checks if $\norm{\boldsymbol{y}-\boldsymbol{\rm {A}}{\boldsymbol{x}}-{b}\cdot(\boldsymbol{\rm {As}}+\boldsymbol{\rm e})} \leq B_P\sqrt{m}$.
    
    \item \textbf{$C=T$}:
    The verifier runs the equation test: the prover is asked to report a bit $c\in \bin$ and a string $d \in \bin^{n\log q}$. The verifier uses the trapdoor $t_k$ to compute the $\boldsymbol{x}_0$ such that $\boldsymbol{y}=\boldsymbol{\rm A}\boldsymbol{x}_0+\boldsymbol{\rm e}_0$, and checks the equation $c=d\cdot({\boldsymbol{x}}_0 \oplus ({\boldsymbol{x}}_0 -{\boldsymbol{\rm s}} ))$ holds.
    \end{enumerate}
\end{enumerate}

 Subsequently, Mahadev et al. in \cite{MVV22} proposed a hypothetical protocol called \emph{quantum verifier lossy randomness protocol} shown in \emph{Protocol 2}, which is essentially similar to the qubit certification protocol, except with the matrix $\boldsymbol{\rm A}$ replaced by an indistinguishable lossy matrix $\boldsymbol{\rm \widetilde{A}}$ and the trapdoor recovery of $\boldsymbol{x}_0$ replaced by a quantum preimage extraction procedure. 

\paragraph{\textbf{Protocol 2: Quantum Verifier Lossy Randomness Protocol}} (Protocol 2.2 in \cite{MVV22})

\begin{enumerate}
  \item The verifier generates $k=(\boldsymbol{\rm \widetilde{A}},\boldsymbol{\rm \widetilde{A}}\boldsymbol{\rm s}+\boldsymbol{\rm e})\in \mathbb{Z}^{m\times n}_q \times \mathbb{Z}^{m}_q$.
  
  \item The verifier sends $k$ to the prover.
  
  \item The prover reports an image ${\boldsymbol{y}}\in \mathbb{Z}^{m}_q$ to the verifier.
  
  \item The verifier chooses a challenge $C\sample \{G,T\}$ and sends $C$ to the prover.
    \begin{enumerate}

    \item \textbf{$C=G$}:
    The verifier runs the preimage test: the prover is asked to return a bit $b$ and a corresponding preimage $\boldsymbol{x}$ of $\boldsymbol{y}$. The verifier checks $\norm{\boldsymbol{y}-\boldsymbol{\rm \widetilde{A}}{\boldsymbol{x}}-{b}\cdot(\boldsymbol{\rm {\widetilde{A}s}}+\boldsymbol{\rm e})} \leq B_P\sqrt{m}$ holds.
    
    \item \textbf{$C=T$}:
    The verifier runs the equation test: the prover is asked to report a bit $c\in \bin$ and a string $d \in \bin^{n\log q}$. The verifier uses the quantum preimage extraction procedure to compute the $\boldsymbol{x}_0$ from $\boldsymbol{y}$, and checks the equation $c=d\cdot({\boldsymbol{x}}_0 \oplus ({\boldsymbol{x}}_0 -{\boldsymbol{\rm s}} ))$ holds.
    \end{enumerate}
\end{enumerate}

 However, this quantum verifier lossy randomness protocol is only be used as an analytical tool. To revert to a classical verifier protocol, the final protocol is a hybrid solution that replaces step iv(b) of \emph{Protocol 2} with step iv.(b) of \emph{Protocol 1}, shown in \emph{Protocol 3}. Note that the key in combining \emph{Protocol 1} and \emph{Protocol 2} is that they are computationally indistinguishable.   

\paragraph{\textbf{Protocol 3: Lossy Randomness Protocol}} (Protocol 2.3 in \cite{MVV22})

\begin{enumerate}
  
  \item The verifier chooses whether to execute a \emph{generation round} ($C=G$) or a \emph{test round} ($C=T$) uniformly at random.
  
  \begin{enumerate}
  \item If $C=G$:
    The verifier samples $k=(\boldsymbol{\rm \widetilde{A}},\boldsymbol{\rm \widetilde{A}}\boldsymbol{\rm s}+\boldsymbol{\rm e})\in \mathbb{Z}^{m\times n}_q \times \mathbb{Z}^{m}_q$. 
    \item If $C=T$:
     The verifier samples $k=(\boldsymbol{\rm A},\boldsymbol{\rm A}\boldsymbol{\rm s}+\boldsymbol{\rm e})\in \mathbb{Z}^{m\times n}_q \times \mathbb{Z}^{m}_q$, along with the trapdoor $t_{\boldsymbol{\rm A}}$. 
     \end{enumerate}
     
    \item The verifier sends $k$ to the prover.
    
    \item The prover reports a ${\boldsymbol{y}}\in \mathbb{Z}^{m}_q$ to the verifier.
    
    \item The verifier chooses a challenge $C\sample \{G,T\}$. 
    \begin{enumerate}
    \item \textbf{$C=G$}:
    The verifier sends $C$ to the prover and receives $(b,\boldsymbol{x})$ from the Prover. Then the verifier checks that $\norm{\boldsymbol{y}-\boldsymbol{\rm \widetilde{A}}{\boldsymbol{x}}-{b}\cdot(\boldsymbol{\rm {\widetilde{A}s}}+\boldsymbol{\rm e})} \leq B_P\sqrt{m}$ holds.
    
    \item \textbf{$C=T$}:
     The verifier sends $C$ to the prover and receives $(c,d)$ from the Prover. The verifier uses the trapdoor $t_k$ to compute the $\boldsymbol{x}_0$ such that $\boldsymbol{y}=\boldsymbol{\rm A}\boldsymbol{x}_0+\boldsymbol{\rm e}_0$, and checks the equation $c=d\cdot({\boldsymbol{x}}_0 \oplus ({\boldsymbol{x}}_0 -{\boldsymbol{\rm s}} ))$ holds.
    \end{enumerate}
\end{enumerate}

 \subsubsection{Our Techniques}
\paragraph{Constructing many-to-one NTCFs.} In the lossy randomness protocol, the $k$-to-1 functions constructed by the leakage resilience play the key role for improving the rate of generation of randomness. Different from \cite{MVV22}, we instead aim to construct a many-to-one trapdoor claw-free functions family based on the original LWE-based NTCF$^1_2$. The techniques are briefly described below.

Firstly, our key point of departure is that whereas a superposition over the claw $\frac{1}{\sqrt{2}}(\ket{0}\ket{\boldsymbol{x}_0}+\ket{1}\ket{\boldsymbol{x}_0-\boldsymbol{\rm s}})$, we consider multiples of $\boldsymbol{\rm s}$ in the original superposition to yield a state of the form $\frac{1}{\sqrt{\kappa}}\sum_b \ket{b}\ket{\boldsymbol{x}_0-b \cdot \boldsymbol{\rm s}}$ for ${b\in \{0,...,\kappa-1\}}$ with $\kappa =\poly$. To achieve this, in analogy to the construction of the NTCF$^1_2$ based on the QLWE assumption, given an LWE instance $(\boldsymbol{\rm A},\boldsymbol{\rm t}=\boldsymbol{\rm{As}}+\boldsymbol{\rm e})$, it is natural to consider defining a NTCF$^1_{\kappa}$ family by letting $f_0(\boldsymbol{x})=\boldsymbol{\rm A}\boldsymbol{x}+\boldsymbol{\rm e}_0, f_1(\boldsymbol{x})=\boldsymbol{\rm A}\boldsymbol{x}+\boldsymbol{\rm e}_0+\boldsymbol{\rm t},\ldots,f_{\kappa-1}(\boldsymbol{x})=\boldsymbol{\rm A}\boldsymbol{x}+\boldsymbol{\rm e}_0+(\kappa-1) \cdot \boldsymbol{\rm t}$. By substituting $\boldsymbol{\rm t}=\boldsymbol{\rm{As}}+\boldsymbol{\rm e}$, it is known that $f_b(\boldsymbol{x})=\boldsymbol{\rm A}(\boldsymbol{x}+b\cdot \boldsymbol{\rm s})+\boldsymbol{\rm e}_0+b\cdot \boldsymbol{\rm e}$ for ${b\in \{0,...,\kappa-1\}}$. Similarly, if $\boldsymbol{\rm e}$ were 0, this would mean that $f_{b}(\boldsymbol{x})=f_{b-1}(\boldsymbol{x}+\boldsymbol{\rm s})=...=f_0(\boldsymbol{x}+b\cdot\boldsymbol{\rm s})$. Indeed, by sampling $\boldsymbol{\rm e}_0$ from a Gaussian much wider than the noisy shifts $b\cdot\boldsymbol{\rm e}$, we also can ensure the distributions $f_b(\boldsymbol{x})$ and $f_0(\boldsymbol{x}+b\cdot\boldsymbol{s})$ are statistically close, thus ensuring that $f_b(\boldsymbol{x})=f_{b-1}(\boldsymbol{x}+\boldsymbol{\rm s})=...=f_0(\boldsymbol{x}+b\cdot\boldsymbol{\rm s})$. We refer to such functions as $\kappa$-to-1 NTCFs. For all claws $(\boldsymbol{x}_0,\boldsymbol{x}_1,...,\boldsymbol{x}_{\kappa-1})$ of the functions, $\boldsymbol{x}_b=(\boldsymbol{x}_0-b\cdot\boldsymbol{\rm s}) \mod q$ for all ${b\in \{0,...,\kappa-1\}}$. 

\begin{figure}[t]
\centering
\includegraphics[width=0.95\textwidth]{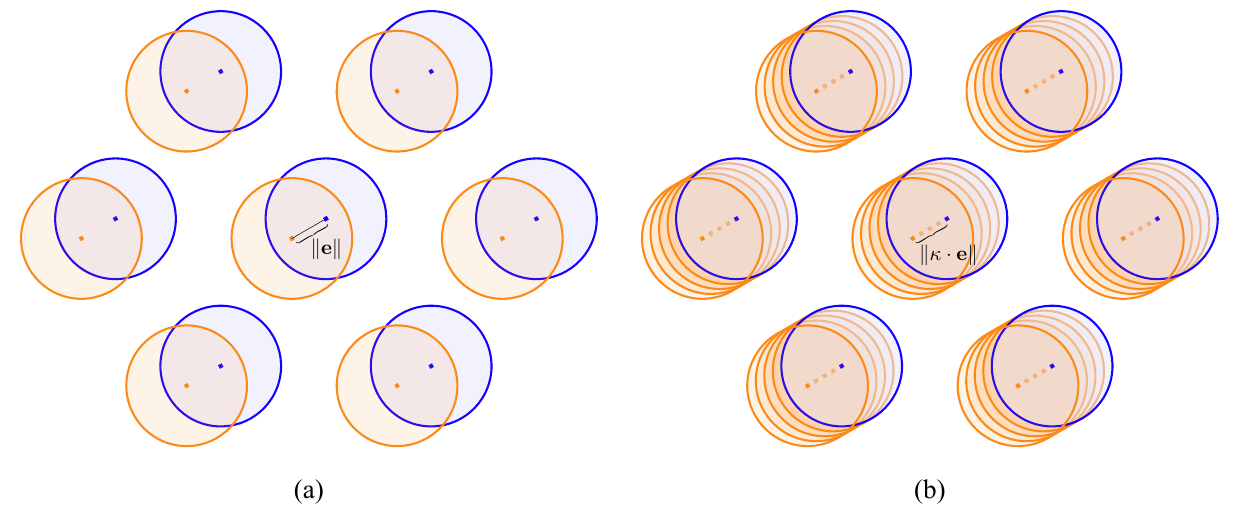}
\caption{The visualization of the balls' intersections technology \cite{BKSW18}. We create spheres with centers lattice points (blue dots) as well as noisy error shifts (orange dots). Once the overlap area is measured, we succeed in outputting a DCP state (as shown in (a)) and an EDCP state (as shown in (b)). In (b), note that the distance between the furthest error shifts and the lattice points $\|\kappa \cdot \boldsymbol{\rm e}\|$ has an upper bound, namely, $\kappa = \poly$ in $n$ dimension lattice.}  \label{fig.3}
\end{figure}

Essentially, the above idea is inspired by the works of the reduction from the LWE to the extrapolated dihedral coset problem (EDCP) in \cite{BKSW18}. Our solution indeed uses the noise flooding technology \cite{Gen09a} and also can be seen as the Balls' intersections method \cite{BKSW18}, as shown in Fig. \ref{fig.3}. 
For the furthest error shifts $\|(\kappa-1)\cdot \boldsymbol{\rm e} \|$, the core is to choose suitable parameter settings such that the two furthest distributions $f_{\kappa-1}(\boldsymbol{x})$ and $f_0(\boldsymbol{x}+(\kappa-1)\cdot\boldsymbol{\rm s})$ are statistically indistinguishable. More details are shown in Section \ref{sec4}. Therefore, leveraging this NTCF$^1_{\kappa}$, the \(\mathsf {QPT}\) prover interacting with the classical verifier according to the qubits certification protocol can easily set up a superposition over $\kappa$ preimages in the form of $\frac{1}{\sqrt{\kappa}}\sum_{b} \ket{b}\ket{(\boldsymbol{x}_0-b \cdot \boldsymbol{\rm s}) \mod{q}}$ for $b\in \{0,...,\kappa-1\}$.

\paragraph{The application for proving quantumness.} Based on the above, we briefly show how to use the NTCF$^1_{\kappa}$ to implement a proof of quantumness protocol. As in \emph{Protocol 1}, in the preimage test, the \(\mathsf {QPT}\) prover can directly measure a superposition over many preimages in the computational basis. However, in the equation test, there is still a left open problem that this superposition over $\kappa$ preimages cannot pass the equation test as it is unable to be checked on the Hadamard basis. To address this issue, we need to introduce a quantum polynomial-time reducing procedure for the \(\mathsf {QPT}\) prover, which transforms a superposition over many preimages into a superposition over two, thereby satisfying the need for the equation test in Hadamard basis \footnote{Note that adding a polynomial-time quantum circuit in the quantum prover's end is an intuitively reasonable assumption, which imposes no additional requirements on the computational power of the verifier and the prover compared to the original protocol in \cite{BCMVV18}.}. The result is then shown in the following Lemma \ref{le_red} and the detail proof is given in Section \ref{sec5.0}.

\begin{lemma} [Informal] \label{le_red}
Let $\mathcal{\widetilde{F}}_{\mathrm{LWE}}$ be a NTCF$^1_{\kappa}$ family based round QLWE assumption. Given a superposition of the form as $\frac{1}{\sqrt{\kappa}}\sum_{b\in \{0,...,\kappa-1\}} \ket{b}\ket{\boldsymbol{x}_0-b \cdot \boldsymbol{\rm s}}$, there exists a quantum reducing procedure $\textsc{RED}_{\mathcal{\widetilde{F}}_{\mathrm{LWE}}}$ consisting of polynomial-time quantum circuits that outputs a superposition of the form  $\frac{1}{\sqrt{2}}(\ket{0}\ket{\bar{\boldsymbol{x}}_0}+\ket{1}\ket{\bar{\boldsymbol{x}}_0-\bar{\boldsymbol{\rm s}} })$ as well as a classical non-zero string $\widehat{b'}$ with probability $\bigOmega{1-1/\poly}$, where $\widehat{b'}=\lvert b-\floor{\frac{\kappa-1}{2}}\rvert$, $\bar{\boldsymbol{x}}_0=(\boldsymbol{x}_0- (\ceil{\frac{\kappa-1}{2}} -\widehat{b'}) \cdot \boldsymbol{\rm s}) \mod{q}$ and $\bar{\boldsymbol{\rm s}}=2\widehat{b'} \cdot \boldsymbol{\rm s} \mod{q}$.
\end{lemma}

Our elementary 2-round proof of quantumness protocol is outlined in \emph{Protocol 4}, which is roughly shown in Fig. \ref{fig.4}. In our protocol, let $\lambda$ be a security parameter and $q\geq 2$ be a prime. Let $\mathcal{\widetilde{F}}_{\mathrm{LWE}}$ be a NTCF$^1_{\kappa}$ family with domain $\mathcal{X}$, range $\mathcal{Y}$ described by the algorithms \textsc{GEN}$_{\mathcal{\widetilde{F}}_{\mathrm{LWE}}}$, \textsc{INV}$_{\mathcal{\widetilde{F}}_{\mathrm{LWE}}}$, \textsc{CHK}$_{\mathcal{\widetilde{F}}_{\mathrm{LWE}}}$ and  \textsc{SAMP}$_{\mathcal{\widetilde{F}}_{\mathrm{LWE}}}$. Let $\textsc{RED}_{\mathcal{\widetilde{F}}_{\mathrm{LWE}}}$ denote a quantum reducing procedure. All setting of parameters is the same as the LWE-based NTCF$^1_{\kappa}$ in Section \ref{sec4.1}. 

\begin{figure}[t]
\begin{tabular}{l|c|l}
\hline
\makecell[c]{\textbf{\(\mathsf {PPT}\) verifier}} & comm. & \makecell[c]{\textbf{\(\mathsf {QPT}\) prover}}\\
\hline
\textbf{Round 1}& & \\
  (1)  $(k,t_k)\leftarrow \textsc{GEN}_{\mathcal{\widetilde{F}}_{\mathrm{LWE}}}(1^{\lambda})$ & $\xrightarrow{~~~k~~~}$ & Get $k=(\boldsymbol{\rm {A}}, \boldsymbol{\rm {As}}+\boldsymbol{\rm e})$\\ [0.8ex]
  & & \makecell[l]{(2) Run $\textsc{SAMP}_{\mathcal{\widetilde{F}}_{\mathrm{LWE}}}$ generate state} \\[0.4ex] 
  & & \makecell[c]{$\sum_{\substack{b,\boldsymbol{x},\boldsymbol{y}}}\sqrt{f'_{k,b}(\boldsymbol{ x})(\boldsymbol{ y})}\ket{b}_{\mathsf B}\ket{\boldsymbol{ x}}_{\mathsf X }\ket{\boldsymbol{ y}}_{\mathsf Y}$} \\[0.8ex] 
  Compute $\boldsymbol{x}_b \gets \textsc{INV}_{\mathcal{\widetilde{F}}_{\mathrm{LWE}}}(t_k,b,\widetilde{\boldsymbol{y}})$ & $\xleftarrow{~~~\widetilde{\boldsymbol{y}}~~~}$ & \makecell[l]{(3) Measure $\mathsf Y$ register, obtain an $\widetilde{\boldsymbol{y}}$ \\and a new state $\sum_{b} \ket{b}_{\mathsf B}\ket{\boldsymbol{x}_b}_{\mathsf X}$}\\
  \hline
  \textbf{Round 2}& & \\
  (5) Choose $C\sample \{G,T\}$ & $\xrightarrow{~~~C~~~}$ & Get $C$\\[0.8ex]
  & & \textbf{If $C=G$} then\\
  \makecell[l]{(7a) Return \emph{Accept} if  $\textsc{CHK}_{\mathcal{\widetilde{F}}_{\mathrm{LWE}}}$ \\ passed}& $\xleftarrow{~~~\widehat b,\widehat{\boldsymbol{x}}_{\widehat b}~~~}$& \makecell[l]{(6a) Measure $\mathsf {BX}$ registers in compu-\\tational basis, return $(\widehat b,\widehat{\boldsymbol{x}}_{\widehat b})$ } \\
   & & \textbf{Else if $C=T$} then\\
  
  \makecell[l]{(7b)Use tuple $(t_k,\boldsymbol{x}_0, \widehat{b'})$ compute \\$\bar{\boldsymbol{x}}_0$ and $\bar{\boldsymbol{x}}_0-\bar{\boldsymbol{\rm s}}$; Return \emph{Accept} if \\$c=d\cdot(\bar{\boldsymbol{x}}_0 \oplus (\bar{\boldsymbol{x}}_0 -\bar{\boldsymbol{\rm s}} ))$ holds  } & $\xleftarrow{~~~\widehat{b'},c,d~~~}$&
  \makecell[l]{(6b) Run $\textsc{RED}_{\mathcal{\widetilde{F}}_{\mathrm{LWE}}}$ get a non-zero $\widehat{b'}$\\ and a state $\ket{0}_{\mathsf B}\ket{\bar{\boldsymbol{x}}_0}_{\mathsf X}+\ket{1}_{\mathsf B}\ket{\bar{\boldsymbol{x}}_0-\bar{\boldsymbol{\rm s}}}_{\mathsf X}$;\\ Measure $\mathsf {BX}$ registers in Hadamard \\basis, return $(c,d)$}\\
  \hline
  
\end{tabular}
\normalsize
\centering
\caption{Schematic representation of our interactive proof of quantumness protocol based on NTCF$^1_{\kappa}$.}
    \label{fig.4}
\end{figure}

\paragraph{\textbf{Protocol 4: The Proof of Quantumness Protocol based on NTCF$^1_{\kappa}$}}
\begin{enumerate}
   \item The verifier generates $(k,t_k)\leftarrow \textsc{GEN}_{\mathcal{\widetilde{F}}_{\mathrm{LWE}}}(1^{\lambda})$, where $k=(\boldsymbol{\rm {A}}, \boldsymbol{\rm {As}}+\boldsymbol{\rm e})$.
  
  \item The verifier sends $k$ to the prover and keeps the trapdoor $t_k$ private.
  
  \item The prover runs $\textsc{SAMP}_{\mathcal{\widetilde{F}}_{\mathrm{LWE}}}$ to return  a state as $\sum_{\substack{b,\boldsymbol{x},\boldsymbol{y}}}\sqrt{f'_{k,b}(\boldsymbol{ x})(\boldsymbol{ y})}\ket{b}\ket{\boldsymbol{ x}}\ket{\boldsymbol{y}}$. After measuring the last register, the prover obtains an image $\widetilde{\boldsymbol{y}}$ and a new state as $\sum_{b} \ket{b}\ket{\boldsymbol{x}_b}$, where $\boldsymbol{x}_b=(\boldsymbol{x}_0-b \cdot \boldsymbol{\rm s}) \mod q$ for $b\in\set{0, \ldots, \kappa -1}$. The prover sends $\widetilde{\boldsymbol{y}}$ to the verifier. For every $b$, the verifier computes $\boldsymbol{x}_b \gets \textsc{INV}_{\mathcal{\widetilde{F}}_{\mathrm{LWE}}}(t_k,b,\widetilde{\boldsymbol{y}})$. 
  
  \item The verifier chooses a challenge $C\sample \{G,T\}$ and sends $C$ to the prover. 
    \begin{enumerate}
    
    \item {$C=G$}:
    The prover returns $\widehat b \in \bin^{\log{\kappa}}$ and a corresponding preimage $\widehat{\boldsymbol{x}}_{\widehat b}\in \mathcal{X}$. The verifier checks that $\norm{\boldsymbol{y}-\boldsymbol{\rm {A}}\widehat{\boldsymbol{x}}_{\widehat b}-\widehat{b}\cdot(\boldsymbol{\rm {As}}+\boldsymbol{\rm e})} \leq B_P\sqrt{m}$ by $\textsc{CHK}_{\mathcal{\widetilde{F}}_{\mathrm{LWE}}}$.
    
    \item \textbf{$C=T$}:
    The prover executes $\textsc{RED}_{\mathcal{\widetilde{F}}_{\mathrm{LWE}}}$ to return a non-zero value $\widehat{b'}$ and a state $\ket{0}\ket{\bar{\boldsymbol{x}}_0}+\ket{1}\ket{\bar{\boldsymbol{x}}_0-\bar{\boldsymbol{\rm s}}}$ (omitting normalization factors), and reports the $\widehat{b'}$ to the verifier. Then the verifier runs an equation test: the verifier uses tuple $(t_k,\boldsymbol{x}_0, \widehat{b'})$ to compute the $\bar{\boldsymbol{x}}_0$ and the $\bar{\boldsymbol{\rm s}}$, and asks the prover for an equation, which consists of $(c,d)\in \bin \times \bin^{n\log q}$. The verifier checks $c=d\cdot(\bar{\boldsymbol{x}}_0 \oplus (\bar{\boldsymbol{x}}_0 -\bar{\boldsymbol{\rm s}} ))$ holds.
    \end{enumerate}
\end{enumerate}

\begin{theorem} [Informal]
Let $\mathcal{\widetilde{F}}_{\mathrm{LWE}}$ be a NTCF$^1_{\kappa}$ family satisfying Definition \ref{Def.m21}. Then the \emph{Protocol 4} satisfies the following properties:
\begin{itemize}
    
    \item[--] Completeness: There exists a \(\mathsf {QPT}\) prover and a negligible function $\negl[\cdot]$ such that for all $\lambda \in \NN$, a \(\mathsf {QPT}\) prover succeeds in the \emph{Protocol 4} with probability at least $1-\negl[\lambda]$.
    \item [--] Soundness: For any classical \(\mathsf {PPT}\) prover, there exists a negligible function $\negl[\cdot]$ such that for all $\lambda \in \NN$, a \(\mathsf {PPT}\) prover succeeds in the \emph{Protocol 4} with probability at most $\negl[\lambda]$.
    
\end{itemize}

\end{theorem}

\subsection{Related Work}\label{sec1.3}

Recently, Leveraging TCFs primitive for quantum cryptographic tasks was pioneered in two breakthrough works: (1) the construction of quantum homomorphic encryption with classical keys \cite{Mah18a} and (2) the verifiable proof of quantumness \cite{BCMVV18}. The latter work also serves as a certifiable quantum randomness generator from an untrusted quantum device. These works have since been extended to other interesting tasks including delegated quantum computation \cite{Mah18b}, quantum money \cite{rad19,Zha21} and remote state preparation \cite{GV19}. Remarkably, instead of LWE-based NTCFs with the adaptive hardcore bit property in \cite{BCMVV18}, recent works accept a broader landscape of TCFs for interactive quantumness tests. Wherein, the construction of TCFs could be based around other hardness assumptions for eliminating the need for an adaptive hardcore bit and using TCFs with a lower circuit complexity, such as Ring-LWE \cite{BKVV20}, learning with rounding (LWR) \cite{LG22}, Rabin's function and Diffie-Hellman problem \cite{KMCVY22}. Recently, Alamati et al. \cite{AMR22} further constructed two-to-one TCFs from isogeny-based group actions. Moreover, there are numerous extensions and follow-up works for TCF-based proofs of quantumness \cite{ACGH20,MV21,BKVV20,MVV22}. Particularly, in \cite{BKVV20}, an efficient non-interactive proof of quantumness protocol was proposed in the quantum random oracle model (QROM) to further reduce the round complexity of \cite{BCMVV18}. In \cite{MVV22}, the work for improving the rate of generation of randomness of \cite{BCMVV18} was proposed, which greatly inspires the present work.

Generally, solving the hidden subgroup problem (HSP) in the dihedral group is unclear even for quantum computation. In \cite{Reg02}, Regev first linked the DHSP to the lattice problems and pointed out that uSVP is no harder to solve than the dihedral coset problem (DCP). Consequently, leveraging the reductions between LWE and unique SVP \cite{LM09,Regev09,AGVW17}, we could naturally bridge the LWE and the DCP. However, the DCP modulus requires super-exponential in $n$ in the reduction of \cite{Reg02}, i.e. $2^{\mathcal{O}(n^2)}$. Recently, Brakerski et al. showed that under quantum polynomial-time reductions, the LWE is equivalent to the \emph{extrapolated dihedral coset problem} (EDCP) \cite{BKSW18}, which can be seen as a special case of the generalized DCP. Besides, since the EDCP is quantumly reducible to the DCP under some certain parameters, the hardness reduction from the LWE to the DCP can be constructed efficiently. Since the work \cite{BKSW18} skips the reduction from BDD to LWE and instead starts from LWE, the modulus of DCP in their reduction is reduced to a polynomial.

To date, finding an efficient solution to DHSP, as well as for DCP and EDCP, remains an open question. The first quantum algorithm for DHSP was proposed by Ettinger and Høyer  \cite{ettinger2000quantum}. They pointed that it is sufficient to solve the DCP by a linear number of coset samples, but post-processing requires exponentially more time. Besides,  Bacon et al. \cite{bacon2005optimal} showed that, no quantum algorithm can find a solution of the DCP with high probability with only a sublinear number of such samples from the DCP oracle. So far the best known quantum algorithm for the DCP is Kuperberg's algorithm \cite{kuper05} which runs in sub-exponential time. Based on this, a series of improvements were subsequently proposed. Regev \cite{regev04subexp} improved Kuperberg's algorithm by using only $poly(\log N)$ space complexity, despite a slight increase in time complexity to $2^{\mathcal{O}(\sqrt{\log N \log\log N})}$. Kuperberg further optimized the space complexity later, whereas the time complexity remains sub-exponential \cite{kuper11}. Recently, inspired by the works of Ivanyos et al. \cite{fri03,fri14,IPS18}, quantum algorithms for solving EDCP via a filtering technique are proposed by Chen et al. \cite{CLZ22}. Notably, the fantastic connection between EDCP and LWE will play the key role in our work.

\subsection{Organization}\label{sec1.5}

The remainder of our paper is organized as follows. Section \ref{sec2} recall the notations and backgrounds on LWE, NTCFs, HSP and DCPs. In Section \ref{sec3}, we show that an interesting inherent connection between LWE-based NTCF$^1_2$ and the dihedral coset problem (DCP). Section \ref{sec4} introduces the definition of NTCF$^1_{\kappa}$ and its construction, and gives a quantum reduction from LWE to the extrapolated dihedral coset problem (EDCP) by using NTCF$^1_{\kappa}$. In Section \ref{sec5}, we describe the same application as \emph{Protocol 1} for a proof of quantumness by using NTCF$^1_{\kappa}$. Finally, the conclusion and open problems are given in Section \ref{Dis}.

\section{Preliminaries}\label{sec2}

\subsection{Notation}\label{sec2.1}

Let $\mathbb{Z}$ is the set of integers, $\mathbb{R}$ is the set of real numbers and $\mathbb{N}$ is the set of natural numbers. For $n\in \mathbb{N}$, let $[n]=\{1,...,n\}$. The vectors are denoted by bold lower case letters (e.g., $\boldsymbol{\rm x}\in \mathbb{Z}^n$), matrices by bold uppercase letters (e.g., $\boldsymbol{\rm A}\in \mathbb{Z}^{m \times n}$). For any $q\in \mathbb{N}$ such that $q \geq 2$, the notation $\mathbb{Z}_q$ is denoted the quotient ring $\mathbb{Z}/q\mathbb{Z}$, i.e., the ring of integers modulo $q$. We write $\negl[\lambda]$ for any function $f:\NN \to \RR_+$ such that for any polynomial $p$, $\lim_{\lambda \to \infty}p(\lambda)f(\lambda)=0$. Let $\poly$ is a polynomial in $n$. Let $\mathsf{PPT}$ stand for classical probabilistic polynomial-time and $\mathsf{QPT}$ stand for quantum polynomial-time.

Let the letter $D$ denote a distribution over a finite domain $X$ and $f$ for a density on $X$, i.e., a function $f:X \to [0,1]$ s.t. $\sum_{x\in X}f(x)=1$. $x\leftarrow D$ indicates that $x$ is sampled from the distribution $D$, and $x \sample X$ indicates that $x$ is sampled uniformly from the set $X$ in random. Let $D_X$ for the set of all densities on $X$. For any $f\in D_X$, $\Supp({f})$ is denoted the support of $f$, $\Supp({f})=\{x\in X| f(x)>0\}$.

For the two densities $f_0,f_1$ over the same domain $X$, the Hellinger distance $H^2(f_0,f_1)$ is defined by 
\begin{align}\label{hellinger}
    H^2(f_0,f_1)=1-\sum_{x\in X}\sqrt{f_0(x)f_1(x)},
\end{align}
the total variation distance between $f_0$ and $f_1$ is given by 
\begin{align}
    \|f_0 - f_1\|_{TV} = \frac{1}{2}\sum_{x \in X} |f_0(x)-f_1(x)|.
\end{align}

\begin{lemma}[Lemma 2.1 in \cite{BCMVV18}]
Let $X$ be a finite set and $f_0,f_1 \in D_X$. Given
\begin{align*}
    \ket{\psi_0}=\sum_{x\in X}\sqrt{f_0(x)}\ket{x},
    \ket{\psi_1}=\sum_{x\in X}\sqrt{f_1(x)}\ket{x},
\end{align*}
then, $\|\ket{\psi_0}-\ket{\psi_1}\|_{tr}=\sqrt{1-(1-H^2(f_0,f_1))^2}$ holds.
\end{lemma}

\begin{definition} [Quantum Fourier Transform (QFT)] \label{QFT}
For any integer $q\geq 2$, let $\omega_q=e^{2\pi i /q}$ denote a primitive $q$-th root of unity. The unitary operator $\textsc{QFT}_q:=F_q \in \CC^{q\times q}$ can be implemented by $\poly[\log q]$ elementary quantum gates, where $(F_q)_{i,j}=\frac{1}{\sqrt{q}}\omega^{i \cdot j}_q$ for $i,j \in \ZZ_q$. The $\textsc{QFT}_q$ is performed in a quantum state $\ket{\phi}:=\sum_{x\in \ZZ_q}f(x)\ket{x}$ yielding 
\begin{align}
    \textsc{QFT}_q \ket{\phi}=\sum_{y\in \ZZ_q} \widehat{f}(y)\ket{y}:=\sum_{y\in \ZZ_q}\sum_{x\in \ZZ_q}\frac{1}{\sqrt{q}}\cdot \omega^{x \cdot y}\cdot f(x) \ket{y}.
\end{align}
\end{definition}


\subsection{The Learning with Errors (LWE) Problem}\label{sec2.2}

For a positive real $B$ and positive integers $m$ and $q$, the truncated discrete Gaussian over $\mathbb{Z}^m_q$ with parameter $B$ is the distribution supported on $\{\boldsymbol{x}\in \mathbb{Z}^m_q:\|\boldsymbol{x}\| \leq B\sqrt{m}\}$ with density
\begin{align} \label{density}
   D_{\mathbb{Z}^m_q,B}(\boldsymbol{x})= D_{\mathbb{Z}_q,B}(x_1)\cdot \cdot \cdot  D_{\mathbb{Z}_q,B}(x_m)
\end{align}
for $\forall \boldsymbol{x}=(x_1,...,x_m)\in \mathbb{Z}^m_q$ and $x_i \in \mathbb{Z}_q$ with $i\in [m]$, where
\begin{align}
    D_{\mathbb{Z}_q,B}(x_i)=\frac{e^{\frac{-\pi |x_i|^2}{B^2}}}{\sum\limits_{x_i\in \mathbb{Z}_q,|x_i|\leq B}e^{{\frac{-\pi |x_i|^2}{B^2}}}}.
\end{align}
\begin{lemma}[Lemma 2.4 in \cite{BCMVV18}] \label{Le2.4}
Let $B$ be a positive real and $q$, $m$ are positive integers. Given $\boldsymbol{\rm e} \in \mathbb{Z}^m_q$ s.t. $\|\boldsymbol{\rm e}\| \leq B\sqrt{m}$. The Hellinger distance between the distribution $D=D_{\mathbb{Z}^m_q,B}$ and the distribution $(D+\boldsymbol{\rm e})$, with density $(D+\boldsymbol{\rm e})(x)=D(x-\boldsymbol{\rm e})$, satisfies
\begin{align} \label{Hellinger}
    H^2(D,D+\boldsymbol{\rm e}) \leq 1- e^{\frac{-2\pi\sqrt{m}\|\boldsymbol{\rm e}\|}{B}},
\end{align}
and the statistical distance between the two distributions satisfies
\begin{align}
    \|D-(D+\boldsymbol{\rm e})\|^2_{TV} \leq 2(1-e^{\frac{-2\pi\sqrt{m}\|\boldsymbol{\rm e}\|}{B}}).
\end{align}
\end{lemma} 

\begin{definition}[Truncated Gaussian superposition, \cite{PA22}] \label{def1}
Let $m\in \mathbb{N}$ and $q\geq 2$ be an integer modulus. Consider $D_{\mathbb{Z}^m_q,B}$ be the truncated discrete Gaussian with parameter $B>0$ over the finite cube $\mathbb{Z}^m \cap (-\frac{q}{2},\frac{q}{2}]^m$. Then the truncated discrete Gaussian superposition state $\ket{D_{\mathbb{Z}^m_q,B}}$ is defined as
    \begin{align}\label{GassState}
        \ket{D_{\mathbb{Z}^m_q,B}}= \sum_{\boldsymbol{x}\in \mathbb{Z}^m_q}\sqrt{D_{\mathbb{Z}^m_q,B}(\boldsymbol{x})}\ket{\boldsymbol{x}}.
    \end{align}
\end{definition}
Note that the truncated Gaussian superposition in Eq.(\ref{GassState}) can be created using the technique via the Grover and Rudolph algorithm \cite{GR02} (see also \cite{BCMVV18}).

    \begin{definition}[Search-LWE]\label{SLWE}
    For a security parameter $\lambda$, let $n,m,q \in \mathbb{N}$ be integer functions of $\lambda$. Let $\chi=\chi(\lambda)$ be a distribution over $\mathbb{Z}_q$. Let $\boldsymbol{s} \in \mathbb{Z}^n_q$ be a fixed row vector chosen uniformly at random. Given access to the LWE samples $\left(\boldsymbol{\rm A},\boldsymbol{\rm {As}}+\boldsymbol{\rm e} \Mod{q}\right) \in \mathbb{Z}^{m\times n}_q \times \mathbb{Z}^m_q$, where $\boldsymbol{\rm A}$ is uniformly random in $\mathbb{Z}^{m\times n}_q$, $\boldsymbol{\rm e}$ is a row vector drawn at random from the distribution $\chi^m$. The search-LWE$_{n,m,q,\chi}$ is to recover $\boldsymbol{\rm s}$. We often denote this problem as search-LWE$_{n,q,\chi}$ for any function $m=\Omega(n\log q)$.
    \end{definition}

    \begin{definition}[Decision-LWE]
    For a security parameter $\lambda$, let $n,m,q \in \mathbb{N}$ be integer functions of $\lambda$. Let $\chi=\chi(\lambda)$ be a distribution over $\mathbb{Z}_q$. Solving the decision-LWE$_n,q,\chi$ is to distinguish between distributions $\left(\boldsymbol{\rm A},\boldsymbol{\rm {As}}+\boldsymbol{\rm e} \Mod{q}\right)$ and $(\boldsymbol{\rm A},\boldsymbol{\rm u})$, where $\boldsymbol{\rm A} \sample \mathbb{Z}^{m\times n}_q$, $\boldsymbol{\rm s}\sample \mathbb{Z}^{n}_q$, $\boldsymbol{\rm e}\leftarrow \chi^m$, and $\boldsymbol{\rm u}\sample \mathbb{Z}^{m}_q$. Likewise, we denote this problem by decision-LWE$_{n,q,\chi}$ for any function $m=\Omega(n\log q)$.
    \end{definition}
    
     Regev proved that given any $\alpha >0$ such that $B=\alpha q \geq 2\sqrt{n}$, the \emph{Gap shortest vector problem} (GapSVP) and the \emph{shortest independent vector problem} (SIVP) are quantumly reducible to the LWE$_{n,q,D_{\mathbb{Z}_q,B}}$ within a factor of $\gamma=\widetilde{O}(n/\alpha)$ in the worst case dimension $n$ lattices \cite{Regev09}.
     

    \begin{theorem}[Theorem 5.1 in \cite{MP12}] \label{Th5.1}
    Let $n,m \geq 1$ and $q \geq 2$ be such that $m=\Omega(n\log q)$. There is an efficient randomized algorithm \textsc{GenTrap}$ (1^n,1^m,q)$ that returns a matrix $\boldsymbol{\rm A} \in \ZZ ^{m\times n}_q$ and a trapdoor $t_{\boldsymbol{\rm A}}$ such that the distribution of $\boldsymbol{\rm A}$ is negligibly (in $n$) close to the uniform distribution. Moreover, there is an efficient algorithm \textsc{Invert} that, on input $\boldsymbol{\rm A},t_{\boldsymbol{\rm A}}$ and $\boldsymbol{\rm A}\boldsymbol{\rm s}+\boldsymbol{\rm e}$ where $\|\boldsymbol{\rm e}\| \leq q/C_T\sqrt{n\log q}$ and $C_T$ is a universal constant, returns $\boldsymbol{\rm s}$ and $\boldsymbol{\rm e}$ with overwhelming probability over $(\boldsymbol{\rm A},t_{\boldsymbol{\rm A}})$ $\gets$ \textsc{GenTrap} $(1^n,1^m,q)$.
    \end{theorem}
    
    \begin{theorem}[Theorem 3.4 in \cite{MVV22}] \label{The3.4}
    Under the QLWE assumption, the distribution of a random $\boldsymbol{\rm \widetilde{A}} \gets \textsc{LOSSY}(1^n,1^m,1^\ell,q,\chi)$ is computationally indistinguishable from $\boldsymbol{\rm A} \sample \ZZ^{m\times n}_q$.
    
    \end{theorem}

\subsection{Noisy Trapdoor Claw-free Functions based on LWE}\label{sec2.3}

Recall from \cite{BCMVV18}, the notion of noisy trapdoor claw-free functions (NTCFs) is introduced and the LWE-based construction of NTCFs is described. 

    \begin{definition}[NTCF family, Definition 3.1 in \cite{BCMVV18}] \label{deft21}
    Let $\lambda$ be a security parameter. Let $\mathcal{X}$ and $\mathcal{Y}$ be finite sets. Let $\mathcal{K_F}$ be a finite set of keys. A family functions $\mathcal{F}=\{ f_{k,b}:\mathcal{X} \to D_y \}_{k \in \mathcal{K_F},b\in \{0,1\}}$ is called an NTCF family if it holds:
    \begin{enumerate}
    \item \textbf {Efficient Function Generation}. There exists an efficient probabilistic algorithm $\textsc{GEN}_{\mathcal{F}}$ which generates a key $k\in \mathcal{K_F}$ together with a trapdoor $t_k$: $(k,t_k)\leftarrow \mathrm{GEN}_{\mathcal{F}}(1^{\lambda})$.
    
    \item \textbf {Trapdoor Injective Pair}. For all $k\in \mathcal{K_F}$, the following conditions hold.
        \begin{enumerate}
            \item Trapdoor: There exists an efficient deterministic algorithm $\textsc{INV}_{\mathcal{F}}$ such that for all $b\in\{0,1\}$, $x\in \mathcal{X}$ and $y\in \Supp (f_{k,b}(x))$, $\textsc{INV}_{\mathcal{F}}(t_k,b,y)=x$. 
            
            \item Injective pair: There exists a perfect matching $\mathcal{R}_k \subseteq \mathcal{X} \times \mathcal{X}$ such that $f_{k,0}(x_0)=f_{k,1}(x_1)$ iff $(x_0,x_1)\in \mathcal{R}_k$.
        \end{enumerate}
    \item \textbf {Efficient Range Superposition}. For all keys $k\in\mathcal{K_F}$ and $b\in \{0,1\}$ there exists a function $f'_{k,b}:\mathcal{X} \to D_\mathcal{Y}$ such that the following hold.
        \begin{enumerate}
            \item for all $(x_0,x_1)\in \mathcal{R}_k$ and $y\in \Supp (f'_{k,b}(x_b))$, $\mathrm{INV}_{\mathcal{F}}(t_k,b,y)=x_b$ and $\mathrm{INV}_{\mathcal{F}}(t_k,b\oplus 1,y)=x_{b\oplus 1}$.
            
            \item There exists an efficient deterministic procedure $\textsc{CHK}_{\mathcal{F}}$ that, on input $k,b\in \{0,1\}$, $x\in \mathcal{X}$ and $y\in \mathcal{Y}$, returns 1 if $y\in \Supp (f'_{k,b}(x))$ and 0 otherwise. Note that $\textsc{CHK}_{\mathcal{F}}$ is not provided the trapdoor $t_k$.
            
            \item For every $k$ and $b\in \{0,1\}$, 
            \begin{align*}
                \expsub{x \sample \mathcal{X}}{H^2(f_{k,b}(x),f'_{k,b}(x))}\leq \negl[\lambda],
            \end{align*}
            where $H^2$ is the Hellinger distance in Eq.(\ref{Hellinger}). Moreover, there exists an efficient procedure $\textsc{SAMP}_{\mathcal{F}}$ that on input $k$ and $b\in\{0,1\}$ creates the state 
            \begin{align*}
                \frac{1}{\sqrt{2|\mathcal{X}|}}\sum_{x\in \mathcal{X},y\in \mathcal{Y},b\in\{0,1\}}\sqrt{(f'_{k,b}(x))(y)}\ket{b,x}\ket{y}.
            \end{align*}
        \end{enumerate}
    \item \textbf {Adaptive Hardcore Bit}. For all keys $k\in \mathcal{K_F}$ the following conditions hold, for some integer $\omega$ that is a polynomially bounded function of $\lambda$.
        \begin{enumerate}
            \item For all $b\in \bin$ and $x\in \mathcal{X}$, there exists a set $G_{k,b,x}\subseteq \bin^{\omega}$ such that $\probsub {d \sample {\bin^{\omega}}}{d \notin G_{k,b,x}}$ is negligible, and moreover there exists an efficient algorithm that checks for membership in $G_{k,b,x}$ given $k,b,x$ and the trapdoor $t_k$.
            
            \item There exists an efficiently computable injection $\mathcal{J}:\mathcal{X} \to \bin^{\omega}$, such that $\mathcal{J}$ can be inverted efficiently on its range, and such that the following holds. If
            \begin{align*}
                \overline{H}_k&= \set{(b,x_b,d,c) \mid b\in \bin, (b,x,d,c\oplus1)\in H_k},\\
                H_k &= \set{(b,x_b,d,d\cdot(\mathcal{J}(x_0)\oplus\mathcal{J}(x_1))) \mid (x_0,x_1)\in \mathcal{R}_k,d\in G_{k,0,x_0} \cap G_{k,1,x_1}}.
            \end{align*}
        \end{enumerate}
    
    For any \(\mathsf {QPT}\) adversary $\mathcal{A}$, there exists a negligible function $\negl[\lambda]$ such that 
    \begin{align*}
       \bigg|\probsub{(k,t_k)\gets{\textsc{GEN}_{\mathcal{F}}(1^\lambda)}}{\mathcal{A}(k)\in H_k}-\probsub{(k,t_k)\gets{\textsc{GEN}_{\mathcal{F}}(1^\lambda)}}{\mathcal{A}(k)\in \overline{H}_k} \bigg| \leq \negl[\lambda].
    \end{align*}
    
    \end{enumerate}
    \end{definition}
    
Next, let us briefly describe the NTCF family $\mathcal{F}_{\mathrm{LWE}}$. Let $\mathcal{X}=\mathbb{Z}^n_q$, and $\mathcal{Y}=\mathbb{Z}^m_q$. Given the key space $\mathcal{K_{\mathcal{F}_{\mathrm{LWE}}}}$ is a subset of $\mathbb{Z}^{m\times n}_q \times \mathbb{Z}^{m}_q$. Indeed, For the key $k=(\boldsymbol{\rm A},\boldsymbol{\rm{As}}+\boldsymbol{\rm e})$, $b\in \{0,1\}$ and $x\in \mathcal{X}$, the ideal density $f_{k,b}(x)$ is defined as
\begin{align}\label{fkb}
    \forall y\in \mathcal{Y}, (f_{k,b}(x))(y)=D_{\mathbb{Z}^m_q,B_P}(y-\boldsymbol{\rm A}x-b \cdot \boldsymbol{\rm{As}}),
\end{align}
 where $D_{\mathbb{Z}^m_q,B_P}$ is the density defined in Eq.(\ref{density}). However, as shown in \cite{BCMVV18}, it is not possible to create Eq.(\ref{fkb}) perfectly in the instantiation based on LWE, but it is possible to create density $(f'_{k,b}(x))(y)$ defined as  
 \begin{align}\label{f'kb}
    \forall y\in \mathcal{Y}, (f'_{k,b}(x))(y)=D_{\mathbb{Z}^m_q,B_P}(y-\boldsymbol{\rm A}x-b \cdot (\boldsymbol{\rm{As}}+\boldsymbol{\rm e})).
\end{align}
Note that for all $x\in \mathcal{X}$, $f'_{k,0}(x)=f_{k,0}(x)$, and the distribution $f'_{k,1}(x)$ and $f_{k,1}(x)$ are shifted by $\boldsymbol{\rm e}$. Indeed, the resulting Eq.(\ref{f'kb}) is within negligible trace distance of Eq.(\ref{fkb}) for a certain choice of parameters shown in Lemma \ref{lemma_NTCF}. Recall the definition of the LWE$_{n,q,\chi}$ defined in Definition \ref{SLWE}, we give the following lemma: 

\begin{lemma}[Theorem 4.1 in \cite{BCMVV18}] \label{lemma_NTCF}
Let $\lambda$ be a security parameter, all other parameters are functions of $\lambda$, $q\geq 2$ be a prime, $n, m \geq 1$ be polynomially bounded functions of $\lambda$, and $B_L, B_V, B_P$ be positive integers. The function family $\mathcal{F}_{\mathrm{LWE}}$ is an NTCF family under the hardness assumption LWE$_{n,q,D_{\mathbb{Z}_q,B_L}}$ if the parameters satisfying the following conditions: $n=\poly[\lambda]$, $m=\Omega(n\log q)$, $B_P=q/2C_T\sqrt{mn\log q}$ where $C_T$ is a universal constant shown in \cite{MP12}, and $B_L<B_V<B_P$ such that the ratios $B_P/B_V$ and $B_V/B_L$ are both super-polynomial in $\lambda$.
\end{lemma}

The possible parameters setting for the secure and correct realization of the construction of $\mathcal{F}_{\mathrm{LWE}}$ has been discussed in \cite{BCMVV18}, where $B_L \geq 2\sqrt{n}$, $q=2^{2\lambda}/\poly[\lambda]$, $\log (q/B_L)=O(\lambda)$ and $B_P/B_V=B_V/B_L=2^{\lambda}$ for aiming $2^{-\lambda}$ statistical security. For ensuring the hardness of LWE$_{n,q,D_{\mathbb{Z}_q,B_L}}$ is exponential in $\lambda$, one can choose $n \approx \Omega(\lambda^2)$ since the problem scales roughly as $2^{\widetilde{\Omega}(n / \log (q/B_L))}$. 

\subsection{Hidden Subgroup Problems}\label{sec3.3}

The hidden subgroup problems are of particular importance in the theory of quantum computation. Many intriguing computational problems can be reduced to an HSP instance, such as integer factoring problem (IFP) and discrete logarithm algorithm (DLP) for Abelian HSP, and lattice problems and graph isomorphism problems for non-Abelian HSP. 

\begin{definition}[Hidden subgroup problem]
The hidden subgroup problem (HSP) is stated that given a function $f:G\to S$, where $G$ is a finite group and $S$ is some finite set, $f$ hides the subgroup $H\leq G$ such that $f(x)=f(y)$ if and only if $x-y\in H$ for all $x, y \in G$. Given the ability to query the function $f$, the goal is to find a generating set for $H$. 
\end{definition}

In particular, for the case of the HSP on Abelian groups, there exists efficient quantum polynomial-time algorithms faster than classical algorithms \cite{shor1994algorithms,simon1997power}, which gives the theoretical evidence for the quantum advantage. The efficient quantum approach for HSP is known as the \emph{standard method}.

\begin{definition} [The standard method, Sect.VII(B) in \cite{childs2010quantum}] \label{SM}
Starting by preparing a uniform superposition over group elements $\ket{G}=\frac{1}{\sqrt{|G|}}\sum_{x\in G}\ket{x}$. Then computing the oracle function $f(x)$ in an ancilla register, yielding the state $\frac{1}{\sqrt{|G|}}\sum_{x\in G}\ket{x,f(x)}$. Note that if we were to measure the second register, obtaining the value $y\in S$, then the state $\frac{1}{\sqrt{|G|}}\sum_{x\in G}\ket{x,f(x)}$ would be projected onto the uniform superposition of those $x\in G$ s.t. $f(x)=y$, which is simply some left coset of $H$. Since every coset has the same number of elements, each left coset occurs with equal probability. Hence, discarding the second register, giving the coset state as
\begin{align}
    \ket{xH}=\frac{1}{\sqrt{|H|}}\sum_{h\in H}\ket{xh},
\end{align}
with $x\in G$ uniformly random and unknown. It is convenient to view the result either as a random pure state, or equivalently, as the mixed state given by $\rho_H = \frac{1}{|G|}\sum_{x\in G}\ket{xH}\bra{xH}$, which is called a hidden subgroup state. In the standard method to the HSP, the goal is to find $H$ using the samples of this hidden subgroup state.
\end{definition}

\subsection{Dihedral Hidden Subgroup Problem}\label{sec3.4}

So far, the ability of quantum computing to solve the HSP on non-Abelian groups remains unclear. As a special case of the non-Abelian groups, the dihedral group $D_N$ is the group of symmetries of an $N$-sided regular polygon, which is generated by the rotation $x$ of angle $2\pi/N$ and reflection $y$ about the $x$-axis, i.e., $D_N$ of order $2N$ is the set $\{x,y|x^N=y^2=yxyx=1_{D_N}\}$. Ettinger and Høyer proved that the dihedral HSP can be reduced to the case where the subgroup $H$ is of the form $\{(0,0),(1,s)\}$ \cite{ettinger2000quantum}. By the \emph{standard method} that samples cosets, the dihedral HSP can be reduced to the dihedral coset problem. 

\begin{definition}[Dihedral hidden subgroup problem]
   The DHSP is the HSP on the dihedral group, $G=D_N \cong \mathbb{Z}_N \rtimes_{\varphi} \mathbb{Z}_2$, where the homomorphism $\varphi$: $\mathbb{Z}_2 \to \mathrm{Aut}(G)$ is defined by $\varphi(0,x)=x$ and $\varphi(1,x)=x^{-1}$. We say that an algorithm solves the DHSP if it runs in (classical or quantum) time $\poly[\log N]$ and succeeds with probability at least $1/{\poly[\log N]}$. 
  \end{definition}

\begin{definition} [Dihedral coset problem (DCP) from \cite{Reg02}]
Generally, the input to the DCP with failure parameter $f\geq 0$ consists of $\poly[\log N]$ registers. Each register is with probability at least $1-\frac{1}{(\log N)^f}$ in the state
\begin{align}\label{DCPst}
   \frac{1}{\sqrt{2}}(\ket{0,x}+\ket{1,(x+s)\mod N}) 
\end{align}
on $1+ \lceil \log N \rceil$ qubits where $x \in \mathbb{Z}_N$ is arbitrary and $s \in \mathbb{Z}_N$ is fixed. Otherwise, with probability at most $\frac{1}{(\log N)^f}$, the bad state is $\ket{b,x}$ where $b\in \{0,1\}$ and $x \in \mathbb{Z}_N$ are arbitrary. We say that an algorithm solves the DCP if it outputs $s$ with probability $\poly[1/(n\log q)]$ in (classical or quantum) time $\poly[n\log q]$.
\end{definition}

\begin{definition} [$\mathbb{Z}^n_q$-dihedral coset problem ($\mathbb{Z}^n_q$-DCP)]\label{ZZ-DCP}
The input to the $\mathbb{Z}^n_q$-DCP with failure parameter $f$ consists of $\poly[n\log q]$ registers. Each register is with probability at least $1-\frac{1}{(n\log q)^f}$ in the state
\begin{align}
  \frac{1}{\sqrt{2}}(\ket{0,\boldsymbol{x}}+\ket{1,(\boldsymbol{x}+\boldsymbol{s}) \mod q})
\end{align}
on $1+ \lceil n\log q \rceil$ qubits where $\boldsymbol{x} \in \mathbb{Z}^n_q$ is arbitrary and $\boldsymbol{s} \in \mathbb{Z}^n_q$ is fixed. Otherwise, with probability at most $\frac{1}{(n\log q)^f}$, the bad state is $\ket{b,\boldsymbol{x}}$ where $b\in \{0,1\}$ and $\boldsymbol{x} \in \mathbb{Z}^n_q$ are arbitrary. We say that an algorithm solves the $\mathbb{Z}^n_q$-DCP if it outputs $\boldsymbol{s}$ with probability $\poly[1/(n\log q)]$ in (classical or quantum) time $\poly[n\log q]$.
\end{definition}

Finally, we introduce a variant of $\mathbb{Z}^n_q$-DCP named the extrapolated dihedral coset problem in \cite{BKSW18}, which is required for the construction of NTCF$^1_{\kappa}$ in Section \ref{sec4}. For convenience, we take $f=0$ for the failure parameter.

\begin{definition} [Extrapolated dihedral coset problem (EDCP)]\label{EDCP}
Let $q\geq 2$ be a modulus, $n\in \mathbb{N}$ be the dimension, and a function $D:\ZZ_q \to \RR$. The input to the $\textsc{EDCP}^\ell_{n,q,D}$ consists of $\ell$ states of the form 
\begin{align}
   \sum_{j\in \Supp(D)}D(j)\ket{j}\ket{(\boldsymbol{x}+j\cdot \boldsymbol{s}) \mod q}, 
\end{align}
where $\boldsymbol{x} \in \mathbb{Z}^n_q$ is arbitrary and $\boldsymbol{s} \in \mathbb{Z}^n_q$ is fixed for all $\ell$ states. We say that an algorithm solves the $\textsc{EDCP}^\ell_{n,q,D}$ if it outputs $\boldsymbol{s}$ with probability $\poly[1/(n\log q)]$ in (classical or quantum) time $\poly[n\log q]$.
\end{definition}

Different choices of $D$ leading to different instantiations of EDCP. There are two interesting cases: (1) $D$ is uniform over $\mathbb{Z}_M$ for some $M\in \mathbb{Z}$, which is denoted as uniform EDCP (U-EDCP$^\ell_{n,q,M}$) and (2) $D$ is Gaussian $D_{\mathbb{Z},B}$, which is denoted as G-EDCP$^\ell_{n,q,B}$. Brakerski et al. showed the G-EDCP and U-EDCP are equivalent up to small parameter losses \cite{BKSW18}. Indeed, the DCP is a special case of EDCP where $n=1$, $q=N$ is exponentially large, and $D$ is the uniform distribution over $\{0,1\}$ \cite{CLZ22}. 

\section{LWE, NTCF$^1_2$ and DCP}\label{sec3}

In this section, we introduce the correlation between the NTCF$^1_2$ based on QLWE and the DCP. Indeed, the NTCF$^1_2$ proposed in \cite{BCMVV18} can be viewed as an explicit description of a DCP oracle function, which can help us to quantumly reduce the LWE to the $\ZZ^n_q$-DCP in polynomial-time.

\subsection{LWE-based NTCF$^1_2$ and Dihedral Coset States} \label{3.1}

We have already introduced the background of the $\ZZ^n_q$-dihedral coset problem ($\ZZ^n_q$-DCP) in the Section \ref{sec3.4}. Here we recall the $\ZZ^n_q$-DCP definition without the failure parameter. (cf. Def. \ref{ZZ-DCP}).

\begin{definition} [$\mathbb{Z}^n_q$-DCP]\label{ZZ-DCP2}
Let $n\in \NN$ be the dimension, $q\geq 2$ be the modulus. The input to the $\mathbb{Z}^n_q$-\textsc{DCP} consists of $\poly[n\log q]$ registers. Each register in the form of
\begin{align}
  \frac{1}{\sqrt{2}}(\ket{0,\boldsymbol{x}}+\ket{1,(\boldsymbol{x}+\boldsymbol{s}) \mod q})
\end{align} \label{nqDCP}
on $1+ \lceil n\log q \rceil$ qubits, where $\boldsymbol{x} \in \mathbb{Z}^n_q$ is arbitrary and $\boldsymbol{s} \in \mathbb{Z}^n_q$ is fixed. We say that an algorithm solves the $\mathbb{Z}^n_q$-DCP if it outputs $\boldsymbol{s}$ with probability $\poly[1/(n\log q)]$ in time $\poly[n\log q]$.
\end{definition}

Next, we will show how to use LWE-based NTCF$^1_2$ to generate a valid sample of $\mathbb{Z}^n_q$-DCP with the secret $\boldsymbol{s} \in \mathbb{Z}^n_q$. Recall from \cite{BCMVV18}, given an LWE$_{n,q,D_{\mathbb{Z}_q,B_V}}$ instance $(\boldsymbol{\rm A},\boldsymbol{t}=\boldsymbol{\rm{As}}+\boldsymbol{\rm e}) \in \mathbb{Z}^{m\times n}_q \times \mathbb{Z}^{m}_q$, where each coordinate of $\boldsymbol{\rm e}$ is drawn independently from distribution $D_{\mathbb{Z}_q,B_V}$,  there exists a polynomial-time quantum procedure SAMP$_{\mathcal{F}_{\mathrm{LWE}}}$ in the construction of the function family $\mathcal{F}_{\mathrm{LWE}}$. 

\begin{lemma}\label{SAM}
    Let $\mathcal{F}_{\mathrm{LWE}}$ be an NTCF$^1_2$ family and $\lambda$ be a security parameter. Let  $\boldsymbol{\rm e} \gets D_{\ZZ^m_q,B_V}$ and $\boldsymbol{\rm e}_0 \gets D_{\ZZ^m_q,B_P}$. For $q=2^{2\lambda}/poly(\lambda)$, $m=\Omega(n\log q)$ and $B_P/B_V=2^{\lambda}$, there exists a quantum polynomial-time procedure $\mathrm{SAMP}_{\mathcal{F}_{\mathrm{LWE}}}$ that on input LWE$_{n,q,D_{\mathbb{Z}_q,B_V}}$ instance $(\boldsymbol{\rm A},\boldsymbol{\rm {As}}+\boldsymbol{\rm e})\in \mathbb{Z}^{m\times n}_q\times \mathbb{Z}^{m}_q$ and $b\in\{0,1\}$, efficiently outputs a superposition 
    \begin{align}
    \frac{1}{\sqrt{2|\mathcal{X}|}}\sum_{x\in \mathcal{X},y\in \mathcal{Y},b\in\{0,1\}}\sqrt{(f'_{k,b}(x))(y)}\ket{b,x}\ket{y},\label{BXY}
    \end{align}
    with probability at least $1-\negl[\lambda]$, where $(f'_{k,b}(x))(y)= {D}_{\mathbb{Z}^m_q,B_P}(y-\boldsymbol{\rm A}x-b \cdot (\boldsymbol{\rm{As}}+\boldsymbol{\rm e}))$ and $k=(\boldsymbol{\rm A},\boldsymbol{\rm {As}}+\boldsymbol{\rm e})$ as shown in Eq.(\ref{f'kb}). 
    \end{lemma}

    \begin{proof}
        Let $\mathcal{X}=\mathbb{Z}^n_q, \mathcal{Y}=\mathbb{Z}^m_q$, the procedure firstly creates the truncated Gaussian superposition state by using the Grover Rudolph technique,
\begin{align}
            \ket{D_{\mathbb{Z}^m_q,B_P}}=\sum_{\boldsymbol{\rm e}_0\in \mathbb{Z}^m_q}\sqrt{D_{\mathbb{Z}^m_q,B_P}(\boldsymbol{\rm e}_0)}\ket{\boldsymbol{\rm e}_0}.
        \end{align}
        Secondly, the procedure creates uniform superposition state over $b\in \{0,1\}$, $x\in \mathcal{X}$ in the form
        \begin{align}
            \frac{1}{\sqrt{2\cdot q^n}}\sum_{\substack{b\in\{0,1\} \\ x\in \mathcal{X}, \boldsymbol{\rm e}_0\in \mathbb{Z}^m_q}}\sqrt{D_{\mathbb{Z}^m_q,B_P}(\boldsymbol{\rm e}_0)}\ket{b,x}\ket{\boldsymbol{\rm e}_0}.
        \end{align}
        Afterward, using the input instance $(\boldsymbol{\rm A},\boldsymbol{\rm{As}}+\boldsymbol{\rm e})$ where each coordinate of $\boldsymbol{\rm e}$ is drawn independently from $D_{\mathbb{Z}_q,B_V}$, the procedure computes
        \begin{align}
            \frac{1}{\sqrt{2\cdot q^n}}\sum_{\substack{b\in\{0,1\} \\ x\in \mathcal{X}, \boldsymbol{\rm e}_0\in \mathbb{Z}^m_q}}\sqrt{D_{\mathbb{Z}^m_q,B_P}(\boldsymbol{\rm e}_0)}\ket{b,x}\ket{\boldsymbol{\rm e}_0}\ket{\boldsymbol{\rm A}x+\boldsymbol{\rm e}_0+b\cdot(\boldsymbol{\rm{As}}+\boldsymbol{\rm e})}.
        \end{align}
        Note that $\boldsymbol{\rm e}_0$ can be computed from $x$, the last register, $b$ and $(\boldsymbol{\rm A},\boldsymbol{\rm{As}}+\boldsymbol{\rm e})$. Then the procedure can uncompute the $\ket{\boldsymbol{\rm e}_0}$ obtaining
        \begin{align}
             \frac{1}{\sqrt{2\cdot q^n}}&\sum_{\substack{b\in\{0,1\} \\ x\in \mathcal{X}, \boldsymbol{\rm e}_0\in \mathbb{Z}^m_q}}\sqrt{D_{\mathbb{Z}^m_q,B_P}(\boldsymbol{\rm e}_0)}\ket{b,x}\ket{\boldsymbol{\rm A}x+\boldsymbol{\rm e}_0+b\cdot(\boldsymbol{\rm {As}}+\boldsymbol{\rm e})}\notag \\
            = \frac{1}{\sqrt{2\cdot q^n}}&\sum_{\substack{b\in\{0,1\} \\ x\in \mathcal{X}, y\in \mathcal{Y}}}\sqrt{D_{\mathbb{Z}^m_q,B_P}(y-\boldsymbol{\rm A}x+b\cdot(\boldsymbol{\rm {As}}+\boldsymbol{\rm e}))}\ket{b,x}\ket{y}\notag \\
            = \frac{1}{\sqrt{2\cdot q^n}}&\sum_{\substack{b\in\{0,1\} \\ x\in \mathcal{X}, y\in \mathcal{Y}}}\sqrt{(f'_{k,b}(x))(y)}\ket{b,x}\ket{y}\notag.
        \end{align} 
    \end{proof}

\begin{lemma}\label{Hnegl}
Let $\lambda$ be a security parameter. Let $\boldsymbol{\rm e} \gets D_{\ZZ^m_q,B_V}$ and $\boldsymbol{\rm e}_0 \gets D_{\ZZ^m_q,B_P}$. For $b\in \bin$, $q=2^{2\lambda}/poly(\lambda)$, $m=\Omega(n\log q)$ and $B_P/B_V=2^{\lambda}$, the Hellinger distance between the distribution $f_{k,b}(x)$ and $f'_{k,b}(x)$ is negligible, where $(f_{k,b}(x))(y)=D_{\mathbb{Z}^m_q,B_P}(y-\boldsymbol{\rm A}x-b \cdot \boldsymbol{\rm{As}})$ and $(f'_{k,b}(x))(y)=D_{\mathbb{Z}^m_q,B_P}(y-\boldsymbol{\rm A}x-b \cdot (\boldsymbol{\rm{As}}+\boldsymbol{\rm e}))$.
\end{lemma}

\begin{proof}
 Since the density $f_{k,b}(x)$ and $f'_{k,b}(x)$ are shifted by $b \cdot \boldsymbol{\rm e}$, using the lemma \ref{Le2.4} we have $H^2(f_{k,0}(x),f'_{k,0}(x)) =0$ and $H^2(f_{k,1}(x),f'_{k,1}(x)) \leq 1-e^{\frac{-2\pi m B_V}{B_P}}$. Based on the assumption that $B_P/B_V$ is super-polynomial, thus $H^2(f_{k,b}(x),f'_{k,b}(x))$  is negligible for every $k$ and $b \in \bin$, namely, $$\expsub{x \sample \mathcal{X}}{H^2(f_{k,b}(x),f'_{k,b}(x))}\leq \negl[\lambda].$$
\end{proof}

Therefore, based on the Lemma \ref{SAM} and the Lemma \ref{Hnegl}, after executing the procedure $\mathrm{SAMP}_{\mathcal{F}_{\mathrm{LWE}}}$ to obtain the superposition in the form of Eq.(\ref{BXY}), we measure the last register of this superposition, thereby resulting a state in the first two registers as
 \begin{align}\label{DCP-}
        \frac{1}{\sqrt{2}}(\ket{0,\boldsymbol{x}_0}+\ket{1,(\boldsymbol{x}_0-{\boldsymbol{\rm s}}) \mod q}),
    \end{align}
where $\boldsymbol{x}_0 \in \mathbb{Z}^n_q$ is arbitrary in domain and $\boldsymbol{\rm s} \in \mathbb{Z}^n_q$ is fixed. Compared to the $\mathbb{Z}^n_q$-DCP state in Eq.(\ref{nqDCP}), for clarity, let $\widetilde{\boldsymbol{\rm s}}= -\boldsymbol{\rm s} \mod q$, then Eq.(\ref{DCP-}) can be rewritten as  
    \begin{align}
        \frac{1}{\sqrt{2}}(\ket{0,\boldsymbol{x}}+\ket{1,(\boldsymbol{x}+\widetilde{\boldsymbol{\rm s}}) \mod q}).
    \end{align}
    Hence, it is not hard to see that this state is essentially a valid instance of the $\mathbb{Z}^n_q$-DCP. 

\subsection{Reducing LWE to $\ZZ^n_q$-DCP via NTCF$^1_2$} \label{sec3.2}

\begin{figure}[t]
\centering
\includegraphics[width=0.95\textwidth]{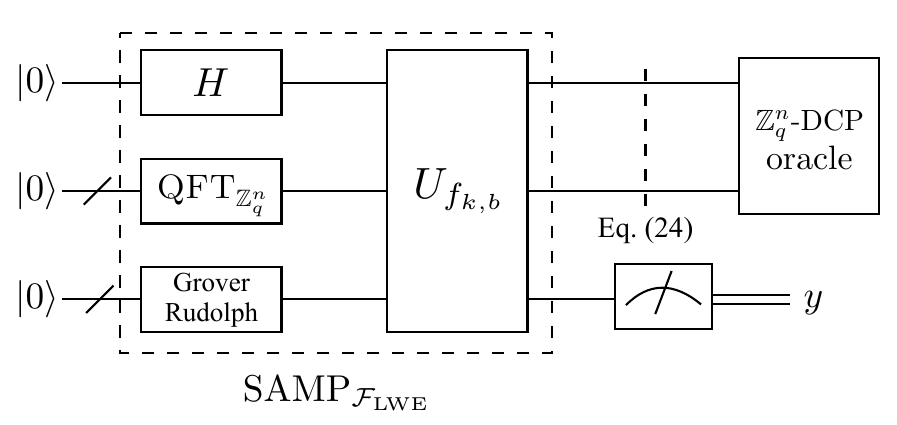}
\caption{Quantum circuit for implementing our reduction LWE $\leq$ $\mathbb{Z}^n_q$-DCP. The input registers are assumed to have the required number of qubits. All global phases are omitted. Given $b\in \bin$ and $k=(\boldsymbol{\rm A},\boldsymbol{\rm {As}}+\boldsymbol{\rm e})$, operation $U_{f_{k,b}}$ is defined as $U_{f_{k,b}}\ket{b}\ket{x}\ket{\boldsymbol{\rm e}_0} \mapsto \ket{b}\ket{x}\ket{\boldsymbol{\rm A}x+\boldsymbol{\rm e}_0+b\cdot(\boldsymbol{\rm {As}}+\boldsymbol{\rm e})\mod q} $.}  \label{fig.5}
\end{figure}

Naturally, our main result in this section is given as follows. The quantum circuit for our quantum reduction LWE$_{n,q,D_{\mathbb{Z}_q,B_V}}$ $\leq$ $\mathbb{Z}^n_q$-DCP is shown in Fig. \ref{fig.5}. 

\begin{theorem} [LWE$_{n,q,D_{\mathbb{Z}_q,B_V}}$ $\leq$ $\mathbb{Z}^n_q$-DCP] \label{theor3.3}
Let $\mathcal{F}_{\mathrm{LWE}}$ be a NTCF$^1_2$ family under the hardness assumption $\mathrm{LWE}_{n,q,D_{\mathbb{Z}_q,B_L}}$ and all parameters are the same as the one in Lemma \ref{lemma_NTCF}. If there exists an efficient quantum algorithm that solves the $\mathbb{Z}^n_q$-DCP, then there exists a quantum solution to the LWE$_{n,q,D_{\mathbb{Z}_q,B_V}}$ problem with super-polynomial modulus.  
\end{theorem}

\begin{proof}
Assume that we are given an LWE$_{n,q,D_{\mathbb{Z}_q,B_V}}$ instance $(\boldsymbol{\rm A},\boldsymbol{\rm {As}}+\boldsymbol{\rm e} \mod q)\in \mathbb{Z}^{m\times n}_q\times \mathbb{Z}^{m}_q$. Our goal is to find $\boldsymbol{\rm s}$ given access to a $\mathbb{Z}^n_q$-DCP. According to the Lemma \ref{SAM} and the Lemma \ref{Hnegl}, we use the quantum procedure $\mathrm{SAMP}_{\mathcal{F}_{\mathrm{LWE}}}$ resulting in the state $\ket{\Psi'}$, where
\begin{align}
   \ket{\Psi'}=\frac{1}{\sqrt{2|\mathcal{X}|}}\sum_{x\in \mathcal{X},y\in \mathcal{Y},b\in\{0,1\}}\sqrt{(f'_{k,b}(x))(y)}\ket{b,x}\ket{y}.
    \end{align}
Since for every $k$ and $b\in \bin$, $\expsub{x \sample \mathcal{X}}{H^2(f_{k,b}(x),f'_{k,b}(x))}\leq \negl[\lambda]$, the statistical distance between $\ket{\Psi'}$ and $\ket{\Psi}$ is at most $\negl[\lambda]$ , where 
\begin{align}
   \ket{\Psi}=\frac{1}{\sqrt{2|\mathcal{X}|}}\sum_{x\in \mathcal{X},y\in \mathcal{Y},b\in\{0,1\}}\sqrt{(f_{k,b}(x))(y)}\ket{b,x}\ket{y}.
    \end{align}
Note that for $\forall y\in \mathcal{Y}$, $(f_{k,0}(x+s))(y)=(f_{k,0}(x))(y)$. Next, after measuring the last register of $\ket{\Psi'}$, the resulting state (ignoring the measure register) with probability $1-\negl[\lambda]$ is
    \begin{align}
        \frac{1}{\sqrt{2}}(\ket{0,\boldsymbol{x}}+\ket{1,(\boldsymbol{x}+\widetilde{\boldsymbol{\rm s}}) \mod q}),
    \end{align}
    where $\boldsymbol{x}\in \ZZ^n_q$ is arbitrary and $\widetilde{\boldsymbol{\rm s}}= -\boldsymbol{\rm s} \mod q$. Thus, repeating the above procedure $\poly[n\log q]$ times, and with probability $(1-\negl[\lambda])^{\poly[n\log q]}$, we obtain $\poly[n\log q]$ many $\mathbb{Z}^n_q$-DCP states. Now we use the $\mathbb{Z}^n_q$-DCP oracle with these states as input and find $\boldsymbol{\rm s}$ as output of the oracle.  
\end{proof}

Remarkably, the $\mathbb{Z}^n_q$-DCP is also quantumly reducible to DCP$_N$ with $N=(2q)^n$ by using the reduction from the \emph{two-point problem} (TPP) to the DCP$_N$ \cite{Reg02}.

\section{LWE, NTCF$^1_{\kappa}$ and EDCP}\label{sec4}

 Following \cite{BKSW18}, since the $\ZZ^n_q$-DCP can be generalized to EDCP, we can analogously extend the NTCF$^1_2$ to the NTCF$^1_{\kappa}$ based around the QLWE assumption. In this section, we first introduce the definition of the NTCF$^1_{\kappa}$ and then present its construction. In addition, we prove that the LWE with super-polynomial modulus can be quantumly reduced to the uniform extrapolated dihedral coset problem (U-EDCP) via the NTCF$^1_{\kappa}$.

\subsection{$\kappa$-to-1 Noisy Trapdoor Claw-free Functions}\label{sec4.0}
Recall from the Definition \ref{EDCP}, inspired by the U-EDCP introduced in \cite{BKSW18}, we extend NTCF$^1_2$ and introduce the notion of NTCF$^1_{\kappa}$ and give a construction based on the QLWE assumption in this subsection. 

\begin{definition}[NTCF$^1_{\kappa}$] \label{Def.m21}
    Let $\mathcal{X}$ and $\mathcal{Y}$ be finite sets. Let $\mathcal{K}_{\mathcal{\widetilde{F}}}$ be a finite set of keys. A family functions
    \begin{align*}
       \mathcal{\widetilde{F}} =\{ f_{k,b}:\mathcal{X} \to D_y \}_{k \in \mathcal{K_{\widetilde{F}}},b\in\set{0, \ldots, \kappa -1}}
    \end{align*}
    is called a $\kappa$-to-1 NTCF family if it holds:
    \begin{enumerate}
    \item \textbf {Efficient Function Generation}. There exists an efficient probabilistic algorithm \textsc{GEN}$_{\mathcal{\widetilde{F}}}$ which generates a key $k\in \mathcal{K_{\widetilde{F}}}$ together with a trapdoor $t_k$: 
        \begin{align*}
            (k,t_k)\leftarrow \textsc{GEN}_{\mathcal{\widetilde{F}}}(1^{\lambda}).
        \end{align*}
    \item \textbf {Trapdoor Injective Pair}. For all keys $k\in \mathcal{K}_{\mathcal{\widetilde{F}}}$, the following conditions hold.
    
        \begin{enumerate}
            \item Trapdoor: There exists a deterministic algorithm \textsc{INV}$_{\mathcal{\widetilde{F}}}$ such that for all $b\in\set{0, \ldots, \kappa -1}$, $x\in \mathcal{X}$ and $y\in \Supp (f_{k,b}(x))$, \textsc{INV}$_{\mathcal{\widetilde{F}}}(t_k,b,y)=x_b$. 
            
            \item Injective pair: There exists a perfect matching $\mathcal{R}_k \subseteq \mathcal{X}^{\kappa}$ such that $f_{k,0}(x_0)=f_{k,1}(x_1)=\ldots =f_{k,\kappa-1}(x_{\kappa-1})$ iff $(x_0,\ldots,x_{\kappa-1})\in \mathcal{R}_k$.
        \end{enumerate}
    
    \item \textbf {Efficient Range Superposition}. For all keys $k\in\mathcal{K_{\widetilde{F}}}$ and $b\in\set{0, \ldots, \kappa-1}$ there exists a function $f'_{k,b}:\mathcal{X} \to D_\mathcal{Y}$ such that the following hold.
        \begin{enumerate}
            \item for all $(x_0,...,x_{\kappa-1})\in \mathcal{R}_k$ and $y\in \Supp (f'_{k,b}(x_b))$, \textsc{INV}$_{\mathcal{\widetilde{F}}}(t_k,b,y)=x_b$ for every $b\in\set{0, \ldots, \kappa -1}$.
            
            \item There exists an efficient deterministic procedure \textsc{CHK}$_{\mathcal{\widetilde{F}}}$ that, on input $k,b\in\set{0, \ldots, \kappa -1}$, $x\in \mathcal{X}$ and $y\in \mathcal{Y}$, returns 1 if $y\in \Supp (f'_{k,b}(x))$ and 0 otherwise. Note that $\mathrm{CHK}_{\mathcal{\widetilde{F}}}$ is not provided the trapdoor $t_k$.
            
            \item For every $k$ and $b\in\set{0, \ldots, \kappa -1}$, 
            \begin{align*}
            \expsub{x \sample \mathcal{X}}{H^2(f_{k,b}(x),f'_{k,b}(x))}\leq \negl[\lambda],
            \end{align*}
            where $H^2$ is the Hellinger distance in Eq.(\ref{Hellinger}). Moreover, there exists an efficient procedure \textsc{SAMP}$_{\mathcal{\widetilde{F}}}$ that on input $k$ and $b\in\set{0, \ldots, \kappa -1}$ creates the state 
            \begin{align*}
                \frac{1}{\sqrt{\kappa|\mathcal{X}|}}\sum_{x\in \mathcal{X},y\in \mathcal{Y},b\in\set{0, \ldots, \kappa -1}}\sqrt{(f'_{k,b}(x))(y)}\ket{b,x}\ket{y}.
            \end{align*}
        \end{enumerate}
        
    \item \textbf {Claw-Free Property}. For any \ppt adversary $\adv$, there exists a negligible function $\negl[\lambda]$ such that 
    \begin{align*}
        \mathrm{Pr} [(x_0,\ldots,x_{\kappa -1})\in \mathcal{R}_k:(k,t_k)\gets \textsc{GEN}_{\mathcal{\widetilde{F}}}(1^{\lambda}), (x_0,\ldots,x_{\kappa -1}) \gets \adv(k)]\leq \negl[\lambda].
    \end{align*}
    \end{enumerate}
    \end{definition}
    
\subsection{The Construction of NTCF$^1_{\kappa}$ Based on LWE}\label{sec4.1}
 Let $\mathcal{\widetilde{F}}_{\mathrm{LWE}}$ be a NTCF$^1_{\kappa}$ family based on the QLWE assumption. For convenience, we first describe the main parameters used in this section. Let $\lambda$ be a security parameter. All other parameters are functions of $\lambda$. Let $q\geq 2$ be a prime, and $\ell, n, m, \kappa \geq 1$ be polynomially bounded functions of $\lambda$. Let $B_L, B_V, B_P$ be positive integers such that the following conditions hold: 

\begin{enumerate}[label=(\roman*)]
    \item $n=\bigOmega{\ell \log q}$ and $m=\bigOmega{n\log q}$,
    \item $\kappa= \poly$,
    \item $B_P=\frac{q}{\kappa \cdot C_T\sqrt{mn\log q}}$, where $C_T$ is a universal constant in Theorem \ref{Th5.1}, 
    \item $2\sqrt{n} \leq B_L<B_V<B_P$ such that the ratios $\frac{B_P}{B_V}$ and $\frac{B_V}{B_L}$ are both super-polynomial in $\lambda$.
\end{enumerate}

 Let the domain is $\mathcal{X}= \ZZ^n_q$ and the range is $\mathcal{Y}= \ZZ^m_q$. The key space $\mathcal{K}_{\mathcal{\widetilde{F}}_{\mathrm{LWE}}}$ is a subset of $\ZZ^{m\times n}_q \times \ZZ^m_q$. Each function key $k=(\boldsymbol{\rm {A}}, \boldsymbol{\rm {As}}+\boldsymbol{\rm e})$. For $b\in\set{0, \ldots, \kappa -1}$, $x\in \mathcal{X}$, $k=(\boldsymbol{\rm {A}}, \boldsymbol{\rm {As}}+\boldsymbol{\rm e})$, the density function $f_{k,b}(x)$ is defined as follows: 
 \begin{align}\label{k-fkb}
    \forall y\in \mathcal{Y}, (f_{k,b}(x))(y)=D_{\mathbb{Z}^m_q,B_P}(y-\boldsymbol{\rm A}x-b \cdot \boldsymbol{\rm{As}}),
\end{align}
 where $D_{\mathbb{Z}^m_q,B_P}$ is the density defined in Eq.(\ref{density}). Here we show that each of the properties of NTCF$^1_{\kappa}$ hold.
 
 \begin{enumerate}
    \item \textbf {Efficient Function Generation}: The key generation algorithm \textsc{GEN}$_{\mathcal{\widetilde{F}}_{\mathrm{LWE}}}$ first samples $(\boldsymbol{\rm A},t_{\boldsymbol{\rm A}})$ $\gets$ \textsc{GenTrap} $(1^n,1^m,q)$ from Theorem \ref{Th5.1}, $\boldsymbol{\rm s} \sample \ZZ^n_q$ and $\boldsymbol{\rm e} \gets D_{\ZZ^m_q,B_V}$. \textsc{GEN}$_{\mathcal{\widetilde{F}}_{\mathrm{LWE}}}$ outputs a key $k=(\boldsymbol{\rm {A}}, \boldsymbol{\rm {As}}+\boldsymbol{\rm e})$ and the trapdoor $t_k=t_{\boldsymbol{\rm {A}}}$.
    
    \item \textbf {Trapdoor Injective Pair}:

        \begin{enumerate}
            \item Trapdoor: For $k=(\boldsymbol{\rm {A}}, \boldsymbol{\rm {As}}+\boldsymbol{\rm e}) \in \mathcal{K}_{\mathcal{\widetilde{F}}_{\mathrm{LWE}}}$, $x\in \mathcal{X}$ and $b\in\set{0, \ldots, \kappa -1}$, the support of $f_{k,b}(x)$ is
            \begin{align}
                \Supp(f_{k,0}(x)) &= \set{\boldsymbol{\rm {A}}x+\boldsymbol{\rm {e}}_0 
                \mid \|\boldsymbol{\rm e}_0\|\leq B_P\sqrt{m}},\\
                \ldots \notag\\
                 \Supp(f_{k,{\kappa-1}}(x)) &= \set{\boldsymbol{\rm {A}}x+(\kappa-1) \cdot \boldsymbol{\rm {As}}+\boldsymbol{\rm {e}}_0 
                \mid \|\boldsymbol{\rm e}_0\|\leq B_P\sqrt{m}}.
            \end{align}
            
            The inversion algorithm \textsc{INV}$_{\mathcal{\widetilde{F}}_{\mathrm{LWE}}}$ takes as input the trapdoor $t_{\boldsymbol{\rm {A}}}$, $b\in\set{0, \ldots, \kappa -1}$ and $y\in \mathcal{Y}$, and returns $\textsc{Invert}(t_{\boldsymbol{\rm {A}}},\boldsymbol{\rm {A}},y)-b\cdot \boldsymbol{\rm {s}}$. From Theorem \ref{Th5.1}, it is known that with overwhelming probability over the choice of $\boldsymbol{\rm {A}}$, for all $y\in \Supp (f_{k,b}(x))$, $\textsc{Invert}(t_{\boldsymbol{\rm {A}}},\boldsymbol{\rm {A}},y)=x-b\cdot \boldsymbol{\rm {s}}$,  \textsc{INV}$_{\mathcal{\widetilde{F}}_{\mathrm{LWE}}}(t_k,b,y)=x_b$. Therefore, it follows that $\textsc{INV}_{\mathcal{\widetilde{F}}_{\mathrm{LWE}}}=x$. 
            
            \item Injective pair: Let $k=(\boldsymbol{\rm {A}}, \boldsymbol{\rm {As}}+\boldsymbol{\rm e})$. Let $\mathcal{R}_k$ be the set of all pairs $(x_0,x_1,\ldots,x_{\kappa-1})$ such that $f_{k,0}(x_0)=f_{k,1}(x_1)=\ldots =f_{k,\kappa-1}(x_{\kappa-1})$. From the construction, this occurs if and only if $x_{b}=x_0-b\cdot \boldsymbol{\rm {s}}$ for all $b\in\set{0, \ldots, \kappa -1}$, i.e. $x_{1}=x_0-\boldsymbol{\rm {s}},...,x_{\kappa-1}=x_0-(\kappa-1)\cdot\boldsymbol{\rm {s}}$.
        \end{enumerate}
    
    \item \textbf {Efficient Range Superposition}. For $k=(\boldsymbol{\rm {A}}, \boldsymbol{\rm {As}}+\boldsymbol{\rm e}) \in \mathcal{K}_{\mathcal{\widetilde{F}}_{\mathrm{LWE}}}$, $x\in \mathcal{X}$ and $b\in\set{0, \ldots, \kappa -1}$, we define the function $f'_{k,b}(x)$ as follows:
    \begin{align}\label{k-f'kb}
    \forall y\in \mathcal{Y}, (f'_{k,b}(x))(y)=D_{\mathbb{Z}^m_q,B_P}(y-\boldsymbol{\rm A}x-b \cdot (\boldsymbol{\rm{As}}+\boldsymbol{\rm e})).
    \end{align}
    
    Note that $f'_{k,0}(x)=f_{k,0}(x)$ for all $x\in \mathcal{X}$. For each $b\in [\kappa-1]$, the distribution $f'_{k,b}(x)$ and $f_{k,b}(x)$ are shifted by $b\cdot\boldsymbol{\rm e}$. Recall that $\boldsymbol{\rm e} \in \ZZ^m_q$ by sampling each coordinate according to the distribution $D_{\ZZ_q,B_V}$ independently. For all $x\in \mathcal{X}$, the densities $f'_{k,0}(x),\ldots,f'_{k,\kappa-1}(x)$ are given by,
     \begin{align}
                \Supp(f'_{k,0}(x)) &= \Supp(f_{k,0}(x)),\\
                \Supp(f'_{k,1}(x)) &= \set{\boldsymbol{\rm {A}}x+\boldsymbol{\rm {e}}_0+\boldsymbol{\rm {As}}+ \boldsymbol{\rm {e}}
                \mid \|\boldsymbol{\rm e}_0\|\leq B_P\sqrt{m}},\\
                \ldots \notag\\
                 \Supp(f'_{k,{\kappa-1}}(x)) &= \set{\boldsymbol{\rm {A}}x+\boldsymbol{\rm {e}}_0+(\kappa-1) \cdot (\boldsymbol{\rm {As}}+\boldsymbol{\rm {e}}) 
                \mid \|\boldsymbol{\rm e}_0\|\leq B_P\sqrt{m}}.
            \end{align}
    
        \begin{enumerate}
            \item Since $B_V < B_P$, it follows that the norm of the maximum noise term $\boldsymbol{\rm {e}}_0+(\kappa-1) \cdot \boldsymbol{\rm {e}}$ is always at most $\kappa\cdot B_P \sqrt{m}$. Hence, the inversion algorithm \textsc{INV}$_{\mathcal{\widetilde{F}}_{\mathrm{LWE}}}$ can be guaranteed to outputs $x$ on input $t_{\boldsymbol{\rm {A}}}$, $b\in\set{0, \ldots, \kappa -1}$ and $y\in \Supp{(f'_{k,b}(x))}$ if we strengthen that $B_P \leq \frac{q}{\kappa \cdot C_T\sqrt{mn\log q}}$. From the Theorem \ref{Th5.1}, this strengthened trapdoor requirement implies that for all $b\in\set{0, \ldots, \kappa -1}$, $(x_0,...,x_{\kappa-1})\in \mathcal{R}_k$ and $y\in \Supp (f'_{k,b}(x_b))$, \textsc{INV}$_{\mathcal{\widetilde{F}}_{\mathrm{LWE}}}(t_{\boldsymbol{\rm A}},b,y)=x_b$.
            
            \item On input $k=(\boldsymbol{\rm {A}}, \boldsymbol{\rm {As}}+\boldsymbol{\rm e}), b\in\set{0, \ldots, \kappa -1}$, $x\in \mathcal{X}$ and $y\in \mathcal{Y}$, the procedure $\textsc{CHK}_{\mathcal{\widetilde{F}}_{\mathrm{LWE}}}$ checks if $\norm{y-\boldsymbol{\rm {A}}x-b\cdot(\boldsymbol{\rm {As}}+\boldsymbol{\rm e})} \leq B_P\sqrt{m}$.   
            
            \item From Lemma \ref{Le2.4}, we bound the Hellinger distance between the densities $f_{k,b}(x)$ and $f'_{k,b}(x)$ with $b\in [\kappa-1]$,
            \begin{align*}\label{H2fk}
             H^2(f_{k,1}(x),f'_{k,1}(x)) &\leq 1-e^{\frac{-2\pi m B_V}{B_P}},\\
             \ldots\\
             H^2(f_{k,\kappa-1}(x),f'_{k,\kappa-1}(x)) &\leq 1-e^{\frac{-2\pi m\cdot (\kappa-1)\cdot B_V}{B_P}}.
             \end{align*}
            
            Since $B_P/B_V$ is super-polynomial and $\kappa= \poly$, for $k=(\boldsymbol{\rm {A}}, \boldsymbol{\rm {As}}+\boldsymbol{\rm e}), b\in\set{0, \ldots, \kappa -1}$, we have $\expsub{x \sample \mathcal{X}}{H^2(f_{k,b}(x),f'_{k,b}(x))}\leq \negl[\lambda]$. Indeed, the definition of  procedure $\textsc{SAMP}_{\mathcal{\widetilde{F}}_{\mathrm{LWE}}}$ is almost identical to the one in \cite{BCMVV18}, only the Hellinger distance need to be bounded by $1-e^{\frac{-2\pi m\cdot (\kappa-1) \cdot B_V}{B_P}}$ rather than $1-e^{\frac{-2\pi m B_V}{B_P}}$. Here we describe the procedure $\textsc{SAMP}_{\mathcal{\widetilde{F}}_{\mathrm{LWE}}}$ in detail. Let $\mathcal{X}=\mathbb{Z}^n_q, \mathcal{Y}=\mathbb{Z}^m_q$, the procedure first creates the truncated Gaussian superposition state by using the Grover Rudolph technique
        \begin{align}
            \ket{D_{\mathbb{Z}^m_q,B_P}}=\sum_{\boldsymbol{\rm e}_0\in \mathbb{Z}^m_q}\sqrt{D_{\mathbb{Z}^m_q,B_P}(\boldsymbol{\rm e}_0)}\ket{\boldsymbol{\rm e}_0}.
        \end{align}
        Secondly, the procedure creates uniform superposition state over $b\in\set{0, \ldots, \kappa -1}$, $x\in \mathcal{X}$ in the form
        \begin{align}
            \frac{1}{\sqrt{\kappa\cdot q^n}}\sum_{\substack{b\in\set{0, \ldots, \kappa -1} \\ x\in \mathcal{X}, \boldsymbol{\rm e}_0\in \mathbb{Z}^m_q}}\sqrt{D_{\mathbb{Z}^m_q,B_P}(\boldsymbol{\rm e}_0)}\ket{b,x}\ket{\boldsymbol{\rm e}_0}.
        \end{align}
        Afterward, using the key $k=(\boldsymbol{\rm A},\boldsymbol{\rm {As}}+\boldsymbol{\rm e})$ and the $b$, the procedure then computes
        \begin{align}
             \ket{\Phi} &=\frac{1}{\sqrt{\kappa\cdot q^n}}\sum_{\substack{b\in\set{0, \ldots, \kappa -1} \\ x\in \mathcal{X}, \boldsymbol{\rm e}_0\in \mathbb{Z}^m_q}}\sqrt{D_{\mathbb{Z}^m_q,B_P}(\boldsymbol{\rm e}_0)}\ket{b,x}\ket{\boldsymbol{\rm A}x+\boldsymbol{\rm e}_0+b\cdot(\boldsymbol{\rm{As}}+\boldsymbol{\rm e})}\notag \\
             &= \frac{1}{\sqrt{\kappa \cdot q^n}}\sum_{\substack{b\in\set{0, \ldots, \kappa -1} \\ x\in \mathcal{X}, y\in \mathcal{Y}}}\sqrt{D_{\mathbb{Z}^m_q,B_P}(y-\boldsymbol{\rm A}x-b\cdot(\boldsymbol{\rm{As}}+\boldsymbol{\rm e}))}\ket{b,x}\ket{y}\notag \\
            &= \frac{1}{\sqrt{\kappa\cdot q^n}}\sum_{\substack{b\in\set{0, \ldots, \kappa -1} \\ x\in \mathcal{X}, y\in \mathcal{Y}}}\sqrt{f'_{k,b}(x)(y)}\ket{b,x}\ket{y} \label{kxy}.
        \end{align}
        \end{enumerate}
        
    \item \textbf {Claw-Free Property}. Suppose that there exists an adversary $\adv$ that, on input $k=(\boldsymbol{\rm A},\boldsymbol{\rm {As}}+\boldsymbol{\rm e})$ can return $(x_0,x_1,\ldots,x_{\kappa-1})\in \mathcal{R}_k$. Then one can use this $\adv$ to break the LWE$_{n,q,D_{\mathbb{Z}_q,B_V}}$ assumption referred to \cite{BCMVV18}, since $x_b-x_0=b\cdot \boldsymbol{\rm s}$ for $b\in [\kappa-1]$.
    \end{enumerate}

\subsection{Reducing LWE to EDCP via NTCF$^1_{\kappa}$}\label{sec4.3}
\begin{figure}[t]
\centering
\includegraphics[width=0.95\textwidth]{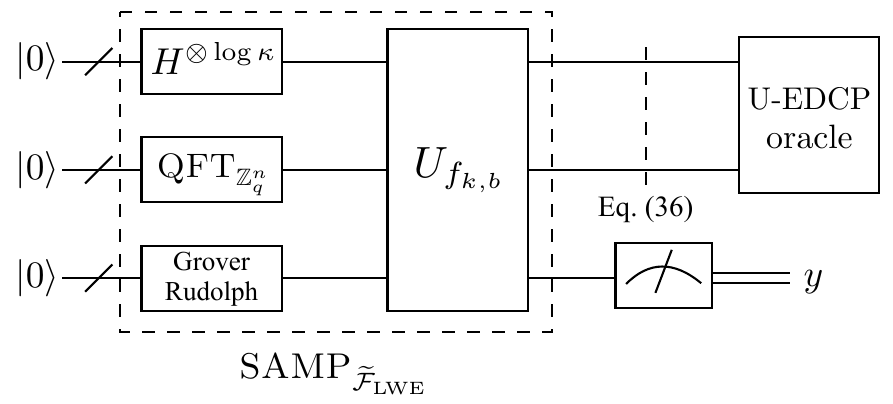}

\caption{Quantum circuit for implementing our reduction LWE $\leq$ U-EDCP. The input registers are assumed to have the required number of qubits. All global phases are omitted. Given $b\in \{0,...,\kappa-1\}$ and $k=(\boldsymbol{\rm A},\boldsymbol{\rm {As}}+\boldsymbol{\rm e})$, operation $U_{f_{k,b}}$ is defined as $U_{f_{k,b}}\ket{b}\ket{x}\ket{\boldsymbol{\rm e}_0} \mapsto \ket{b}\ket{x}\ket{\boldsymbol{\rm A}x+\boldsymbol{\rm e}_0+b\cdot(\boldsymbol{\rm {As}}+\boldsymbol{\rm e})\mod q} $.}  \label{fig.6}
\end{figure}

We have proved that the LWE is reducible to the $\ZZ^n_q$-DCP in Section \ref{sec3.2}. Analogously, here we further prove that there exists a quantum polynomial-time reduction from LWE to U-EDCP with the same parameter settings of the LWE-based NTCF$^1_{\kappa}$. The quantum circuit for our quantum reduction LWE$_{n,q,D_{\mathbb{Z}_q,B_V}}$ $\leq$ U-EDCP$^\ell_{n,q,\kappa}$ is shown in Fig. \ref{fig.6}.

\begin{theorem}[LWE$_{n,q,D_{\mathbb{Z}_q,B_V}}$ $\leq$ U-EDCP$^\ell_{n,q,\kappa}$] \label{theor4.1}
Let $\mathcal{\widetilde{F}}_{\mathrm{LWE}}$ be a NTCF$^1_{\kappa}$ family under the hardness assumption $\mathrm{LWE}_{n,q,D_{\mathbb{Z}_q,B_L}}$ and all parameters are the same as the $\mathcal{\widetilde{F}}_{\mathrm{LWE}}$ shown in section \ref{sec4.1}. If there exists a polynomial time algorithm that solves the U-EDCP$^\ell_{n,q,\kappa}$, then there exists a quantum polynomial-time algorithm that solves the $\mathrm{LWE}_{n,q,D_{\mathbb{Z}_q,B_V}}$, with modulus $q$ is super-polynomial in $\lambda$. 
\end{theorem}

\begin{proof}
    Given an LWE$_{n,q,D_{\mathbb{Z}_q,B_V}}$ instance $(\boldsymbol{\rm A},\boldsymbol{\rm {As}}+\boldsymbol{\rm e} \mod q)\in \mathbb{Z}^{m\times n}_q\times \mathbb{Z}^{m}_q$, the goal is to find $\boldsymbol{\rm s}$ given access to an U-EDCP oracle. According to Section \ref{sec4.1}, let $\mathcal{X}=\ZZ^n_q$, $\mathcal{Y}=\ZZ^m_q$ and $\kappa=\poly$, on input LWE$_{n,q,D_{\mathbb{Z}_q,B_V}}$ instance $(\boldsymbol{\rm A},\boldsymbol{\rm {As}}+\boldsymbol{\rm e} \mod q)$ and $b\in\{0,...,\kappa-1\}$, we can apply the efficient procedure  $\textsc{SAMP}_{\mathcal{\widetilde{F}}_{\mathrm{LWE}}}$ in superposition yielding the state 
\begin{align}
    \frac{1}{\sqrt{\kappa\cdot q^n}}\sum_{\substack{b\in\set{0, \ldots, \kappa -1} \\ \boldsymbol{ x}\in \ZZ^n_q, \boldsymbol{ y}\in \ZZ^m_q}}\sqrt{f'_{k,b}(\boldsymbol{ x})(\boldsymbol{ y})}\ket{b,\boldsymbol{ x}}\ket{\boldsymbol{ y}},
\end{align}
    After measuring the last register to obtain an $\boldsymbol{y}\in\ZZ^m_q$, depending on the definition of NTCF$^1_{\kappa}$, the renormalized state in the first two registers is 
 \begin{align}
     \frac{1}{\sqrt{\kappa}}\sum_{b\in \{0,...,\kappa-1\}} \ket{b}\ket{(\boldsymbol{x}_0-b \cdot \boldsymbol{\rm s}) \mod{q}}\label{ecs},
 \end{align}
    with the success probability of $1-\negl[\lambda]$. We repeat the above procedure $\ell$ times, and with probability $(1-\negl[\lambda])^{\ell}$, we can obtain $\ell$ many  U-EDCP$^{\ell}_{n,q,\kappa}$ states of the form
    \begin{align}
        \left\{ \sum_{ b\in \{0,...,\kappa-1\}} \ket{b}\ket{(\boldsymbol{x}^{(i)}_0-b \cdot \boldsymbol{\rm s}) \mod{q}} \right\} _{i \in [\ell]},
    \end{align}
    where $\boldsymbol{x}^{(i)}_0 \in \ZZ^n_q$ is $i$-th sample value of $\boldsymbol{x}_0$. Now we can call the U-EDCP$^{\ell}_{n,q,\kappa}$ oracle with the above states as input and obtain $\boldsymbol{\rm s}$ as output of oracle.  
    
\end{proof}

Furthermore, recall the EDCP is equivalent to the LWE proved in \cite{BKSW18}. Hence, leveraging our results described in Theorem \ref{theor3.3} and Theorem \ref{theor4.1}, it is not hard to see that the following corollary holds.
\begin{corollary}[U-EDCP $\leq$ $\mathbb{Z}^n_q$-DCP, Lemma 11 \cite{BKSW18}]
If there exists an algorithm that solves the $\mathbb{Z}^n_q$-DCP, then there exists a quantum algorithm for the U-EDCP$^\ell_{n,q,\kappa}$.
\end{corollary}

\section{Application}\label{sec5}

In analogy to the work of \cite{BCMVV18}, our formal protocol consists in the sequential repetition of the 2-round \emph{Protocol 4} between a $\mathsf{PPT}$ verifier and an untrusted $\mathsf{QPT}$ prover, which is outlined in Fig. \ref{fig.7}. Note that to boil down our protocol to the \emph{Protocol 1}, in each equation test round, our protocol requires that the $\mathsf{QPT}$ prover to perform an additional polynomial-time post-processing procedure, which is described in subsection \ref{sec5.0}.   

\begin{figure*}[htbp]
 \begin{framed}  
   \centerline{\textbf{Our protocol for testing quantumness}}
  \vspace{10pt}
  Let $\lambda$ be a security parameter and $q\geq 2$ be a prime. Let $\mathcal{\widetilde{F}}_{\mathrm{LWE}}$ be a NTCF$^1_{\kappa}$ family of functions described by the algorithms \textsc{GEN}$_{\mathcal{\widetilde{F}}_{\mathrm{LWE}}}$, \textsc{INV}$_{\mathcal{\widetilde{F}}_{\mathrm{LWE}}}$, \textsc{CHK}$_{\mathcal{\widetilde{F}}_{\mathrm{LWE}}}$ and \textsc{SAMP}$_{\mathcal{\widetilde{F}}_{\mathrm{LWE}}}$. Let $\textsc{RED}_{\mathcal{\widetilde{F}}_{\mathrm{LWE}}}$ denote the quantum reducing procedure. Let $\ell,n,m,\kappa$ be polynomially bounded functions of $\lambda$, and $B_L,B_V,B_P$ be positive integers. Let $\boldsymbol{\rm e} \gets D_{\ZZ^m_q,B_V}$ and $\boldsymbol{\rm e}_0 \gets D_{\ZZ^m_q,B_P}$. All parameters satisfy the conditions of $\mathcal{\widetilde{F}}_{\mathrm{LWE}}$ in Section \ref{sec4.1}. The $\mathsf{PPT}$ verifier and the $\mathsf{QPT}$ prover repeat the following steps $N$ times: 
\begin{enumerate}

  \item The verifier samples a NTCF$^1_{\kappa}$ key $(k,t_k)\leftarrow \textsc{GEN}_{\mathcal{\widetilde{F}}_{\mathrm{LWE}}}(1^{\lambda})$, where $k=(\boldsymbol{\rm {A}}, \boldsymbol{\rm {As}}+\boldsymbol{\rm e})$.
  
  \item The verifier sends $k$ to the prover and keeps the trapdoor $t_k$ private.
  
  \item The prover uses $k$ to run $\textsc{SAMP}_{\mathcal{\widetilde{F}}_{\mathrm{LWE}}}$ and returns an image $\widetilde{\boldsymbol{y}}\in \mathcal{Y}$ and the superposition state:
  \begin{align*}
     \frac{1}{\sqrt{\kappa}}\sum_{b\in \{0,...,\kappa-1\}} \ket{b}\ket{(\boldsymbol{x}_0-b \cdot \boldsymbol{\rm s}) \mod{q}}.
 \end{align*}
  The prover sends $\widetilde{\boldsymbol{y}}$ to the verifier. For $b\in\set{0, \ldots, \kappa -1}$, the verifier computes $\boldsymbol{x}_b \gets \textsc{INV}_{\mathcal{\widetilde{F}}_{\mathrm{LWE}}}(t_k,b,\widetilde{\boldsymbol{y}})$. 
  
  \item The verifier chooses a uniformly random challenge $C \sample \{G,T\}$ and sends $C$ to the prover. 
    \begin{enumerate}
    
    \item \textbf{Generation round $(C=G)$}:
    The prover returns a string $\widehat b \in \bin^{\log{\kappa}}$ and a corresponding preimage $\widehat{\boldsymbol{x}}_{\widehat b}\in \mathcal{X}$. The verifier checks that $\norm{\boldsymbol{y}-\boldsymbol{\rm {A}}\widehat{\boldsymbol{x}}_{\widehat b}-\widehat{b}\cdot(\boldsymbol{\rm {As}}+\boldsymbol{\rm e})} \leq B_P\sqrt{m}$ by $\textsc{CHK}_{\mathcal{\widetilde{F}}_{\mathrm{LWE}}}$.
    
    \item \textbf{Test round $(C=T)$}:
    The prover executes $\textsc{RED}_{\mathcal{\widetilde{F}}_{\mathrm{LWE}}}$ to return a non-zero value $\widehat{b'}$ and a state $\ket{0}\ket{\bar{\boldsymbol{x}}_0}+\ket{1}\ket{\bar{\boldsymbol{x}}_0-\bar{\boldsymbol{\rm s}}}$, and reports the $\widehat{b'}$ to the verifier. Then the verifier runs an equation test: the verifier uses $\boldsymbol{x}_0$, $\widehat{b'}$ and $t_k$ to compute the $\bar{\boldsymbol{x}}_0$ and the $\bar{\boldsymbol{\rm s}}$, and asks the prover for an equation, which consists of $(c,d)\in \bin \times \bin^{n\log q}$. The verifier checks $c=d\cdot(\bar{\boldsymbol{x}}_0 \oplus (\bar{\boldsymbol{x}}_0 -\bar{\boldsymbol{\rm s}} ))$ holds.
    \end{enumerate}
\end{enumerate}
At the end of the $N$ rounds, if the verifier has not aborted, it will accept.
  
 \end{framed}
 
 \centering
 \caption{The proof of quantumness protocol based on NTCF$^1_{\kappa}$.}
 \label{fig.7}
\end{figure*}

\subsection{Quantum Polynomial-time Reducing Procedure} \label{sec5.0}

\begin{lemma}\label{e2d}
For any $\lambda$, $\kappa\in \poly$ and $b\in\set{0, \ldots, \kappa -1}$, there exists a quantum reducing procedure $\textsc{RED}_{\mathcal{\widetilde{F}}_{\mathrm{LWE}}}$ with run-time polynomial in $\lambda$ that, takes as input state in the form of $\frac{1}{\sqrt{\kappa}}\sum_{b\in \{0,...,\kappa-1\}} \ket{b}\ket{(\boldsymbol{x}_0-b \cdot \boldsymbol{\rm s}) \mod{q}}$, outputs a non-zero value $\widehat{b'}$ and a state
\begin{align}
     \frac{1}{\sqrt{2}}(\ket{0}\ket{\bar{\boldsymbol{x}}_0}+\ket{1}\ket{(\bar{\boldsymbol{x}}_0-\bar{\boldsymbol{\rm s}}) \mod{q}})
\end{align}
with probability $\bigOmega{1-1/\poly}$, where $\widehat{b'}=\lvert b-\floor{\frac{\kappa-1}{2}}\rvert$, $\bar{\boldsymbol{x}}_0=(\boldsymbol{x}_0 - (\ceil{\frac{\kappa-1}{2}} -\widehat{b'}) \cdot \boldsymbol{\rm s}) \mod{q}$ and $\bar{\boldsymbol{\rm s}}=2\widehat{b'} \cdot \boldsymbol{\rm s} \mod{q}$.
\end{lemma}

\begin{proof}
Refer to the Lemma 9 and the Lemma 11 in \cite{BKSW18}, for the uniform EDCP state in the form of Eq.(\ref{ecs}), symmetrizing the uniform distribution by performing the function $f(x)=x-\floor{(\kappa-1)/2}$ to the first register, 
\begin{align*}
     \ket{\psi} &= \frac{1}{\sqrt{\kappa}}\sum_{b\in [0,\kappa-1]} \ket{b-\floor{(\kappa-1)/2}}\ket{(\boldsymbol{x}_0-b \cdot \boldsymbol{\rm s}) \mod{q}}\\
     &= \frac{1}{\sqrt{\kappa}}\sum_{b'\in [-\lfloor \frac{\kappa-1}{2} \rfloor,\lceil \frac{\kappa-1}{2} \rceil]} \ket{b'}\ket{(\boldsymbol{x}'_0-b' \cdot \boldsymbol{\rm s}) \mod{q}},
 \end{align*}
 where $b'=b-\lfloor \frac{\kappa-1}{2} \rfloor$, $\boldsymbol{x}'_0=\boldsymbol{x}_0 - \lceil \frac{\kappa-1}{2} \rceil \cdot \boldsymbol{\rm s}$. 
 Afterward, computing the absolute value function $f(b')=|b'|$ and store it in an ancilla register, yielding the state
 \begin{align*}
     \frac{1}{\sqrt{\kappa}}\sum_{b'\in [-\lfloor \frac{\kappa-1}{2} \rfloor,\lceil \frac{\kappa-1}{2} \rceil]} \ket{b'}\ket{(\boldsymbol{x}'_0-b' \cdot \boldsymbol{\rm s}) \mod{q}}\ket{|b'|}.
 \end{align*}
 Now, measuring the last register and obtaining a value denoted by $\widehat{b'}$. If the observed value $\widehat{b'}$ is zero, we discard it and restart the procedure. For the $\widehat{b'}\neq 0$ with probability $\bigOmega{1-1/\poly}$, then we discard the last register and produce the state as 
 \begin{align*}
     \frac{1}{\sqrt{2}} |-\widehat{b'}\rangle|(\boldsymbol{x}'_0+\widehat{b'} \cdot \boldsymbol{\rm s}) \mod{q}\rangle+\frac{1}{\sqrt{2}}|\widehat{b'}\rangle|(\boldsymbol{x}'_0-\widehat{b'} \cdot \boldsymbol{\rm s}) \mod{q}\rangle.
 \end{align*}
 Since the value $\widehat{b'}$ is known classically, we can uncompute the first register and obtain a DCP state as 
 \begin{align*}
     \frac{1}{\sqrt{2}}(\ket{0}\ket{\bar{\boldsymbol{x}}_0}+\ket{1}\ket{(\bar{\boldsymbol{x}}_0-\bar{\boldsymbol{\rm s}}) \mod{q}}),
 \end{align*}
 where $\bar{\boldsymbol{x}}_0=(\boldsymbol{x}_0'+\widehat{b'} \cdot \boldsymbol{\rm s}) \mod{q}$ and $\bar{\boldsymbol{\rm s}}=2\widehat{b'} \cdot \boldsymbol{\rm s} \mod{q}$.
   
\end{proof}

\subsection{Analysis}\label{5.1} 

\subsubsection{Completeness}\label{5.1.1} 
Here we describe the intended behavior for the \(\mathsf {QPT}\) prover in our protocol. Let an NTCF$^1_{\kappa}$ family $\mathcal{\widetilde{F}}_{\mathrm{LWE}}$ and a key $k\in \mathcal{K}_{\mathcal{\widetilde{F}}_{\mathrm{LWE}}}$. In each round $N_i$ for $i=1,\ldots,N$, an \emph{honest} prover runs the following actions.  

Firstly, the prover can execute the efficient procedure $\textsc{SAMP}_{\mathcal{\widetilde{F}}_{\mathrm{LWE}}}$ in superposition yielding the state 
\begin{align*}
   |\Psi^{(1)}\rangle= \frac{1}{\sqrt{\kappa\cdot q^n}}\sum_{\substack{b\in\set{0, \ldots, \kappa -1} \\ \boldsymbol{ x}\in \ZZ^n_q, \boldsymbol{ y}\in \ZZ^m_q}}\sqrt{f'_{k,b}(\boldsymbol{ x})(\boldsymbol{ y})}\ket{b,\boldsymbol{ x}}\ket{\boldsymbol{ y}}.
\end{align*}

Afterward, the prover measures the last register to obtain an $\boldsymbol{y}\in\ZZ^m_q$, depending on the definition of NTCF$^1_{\kappa}$, the renormalized state in the first two registers is 
 \begin{align*}
     |\Psi^{(2)}\rangle=\frac{1}{\sqrt{\kappa}}\sum_{b\in \{0,...,\kappa-1\}} \ket{b}\ket{(\boldsymbol{x}_0-b \cdot \boldsymbol{\rm s}) \mod{q}},
 \end{align*}
where for $b\in \{0,...,\kappa-1\}$, $\boldsymbol{x}_b = \textsc{INV}_{\mathcal{\widetilde{F}}_{\mathrm{LWE}}}(t_k,b,\widetilde{\boldsymbol{y}})$. Then, according to the verifier's challenge:  

\begin{enumerate}[label=\arabic*)]
    \item In $C_i=T$: the prover executes the procedure $\textsc{RED}_{\mathcal{\widetilde{F}}_{\mathrm{LWE}}}$ to obtain the state 
    \begin{align*}
        |\Psi^{(3)}\rangle=\frac{1}{\sqrt{2}}(\ket{0}\ket{\bar{\boldsymbol{x}}_0}+\ket{1}\ket{\bar{\boldsymbol{x}}_1})|\widehat{b'}\rangle,
    \end{align*}
    where $\bar{\boldsymbol{x}}_1=(\bar{\boldsymbol{x}}_0-\bar{\boldsymbol{\rm s}}) \mod{q}$, $\bar{\boldsymbol{x}}_0=(\boldsymbol{x}_0'+\widehat{b'} \cdot \boldsymbol{\rm s}) \mod{q}$ and $\bar{\boldsymbol{\rm s}}=2\widehat{b'} \cdot \boldsymbol{\rm s} \mod{q}$. Next, the prover evaluates the function $\mathcal{J}(\cdot)$ onto the second register and then applies a Hadamard transform in the first two registers. Tracing out the last register, yielding the state (omitting normalization factors)
    \begin{align*}
        |\Psi^{(4)}\rangle=\sum_{d\in\bin^{n\log q}} (-1)^{d\cdot\mathcal{J}(\bar{\boldsymbol{x}}_0)} \ket{d\cdot (\mathcal{J}(\bar{\boldsymbol{x}}_0)\oplus\mathcal{J}(\bar{\boldsymbol{x}}_1))}\ket{d}.
    \end{align*}
    The prover measures state $|\Psi^{(4)}\rangle$ in computational basis to obtain an $(c,d)$, where $c=d\cdot (\mathcal{J}(\bar{\boldsymbol{x}}_0)\oplus\mathcal{J}(\bar{\boldsymbol{x}}_1))$. The Prover sends $(c,d)$ to the verifier.
    
    \item In $C_i=G$: the prover measures the first two registers of the state $|\Psi^{(2)}\rangle$ in a computational basis, and returns the outcome $(\widehat b,\widehat{\boldsymbol{x}}_{\widehat b})$ to the verifier.  
\end{enumerate}

\begin{lemma}
For any $\lambda$ and $k\gets \textsc{GEN}_{\mathcal{\widetilde{F}}_{\mathrm{LWE}}}$, the honest \(\mathsf {QPT}\) prover in one round of our protocol (with non-zero $\widehat{b'}$ in test round) can be implemented in time polynomial in $\lambda$ and is accepted with probability $1-\negl[\lambda]$.

\end{lemma}

\begin{proof}
    The correctness and efficiency of the \(\mathsf {QPT}\) prover follow from the definition of the NTCF$^1_{\kappa}$ (Definition \ref{Def.m21}). The prover fails only if it obtains an outcome $\widehat{b'} = 0$ shown in Lemma \ref{e2d}, which happens with $1/\poly$. Otherwise, the strategy for honest prover in protocol 4 with non-zero $\widehat{b'}$ is accepted with probability negligibly close to 1. 
    
\end{proof}

\subsubsection{Proof of Quantumness}\label{5.1.2} 

Since the final equation test of our protocol is consistent with the original qubit certification protocol (refer to \emph{Protocol 1}) showed in \cite{BCMVV18}, our proof of quantumness protocol against \(\mathsf {PPT}\) provers is also consistent with the \emph{Protocol 1}, which is shown in Lemma \ref{lemma1}.    

\begin{lemma} \label{lemma1}
 Assuming the QLWE assumption, let $\mathcal{\widetilde{F}}_{\mathrm{LWE}}$ be a family of NTCF$^1_{\kappa}$ satisfying Definition \ref{Def.m21}. For any classical \(\mathsf {PPT}\) prover, in our protocol shown in Fig. \ref{fig.7}, it is the case that $p_{G}+2p_{T}-2\leq \negl[\lambda]$, where $p_{G}$ is  \(\mathsf {PPT}\) prover's success probability in the preimages test round ($C=G$) and $p_{T}$ is \(\mathsf {PPT}\) prover's success probability in the equation test round ($C=T$).
\end{lemma}

\begin{proof}
We aim to find the probability that the \(\mathsf {PPT}\) prover both responds correctly for preimages test round $(C=G)$ and, for the same image $\widetilde{\boldsymbol{y}}$ that they committed to the verifier, equation test round $(C=T)$ is also correct with probability greater than $1/2 + \negl[\lambda]$. Denote this second probability as
\begin{equation*}
    p_{\mathrm{good}} \equiv \Pr_{\widetilde{\boldsymbol{y}}} [p_{T,\widetilde{\boldsymbol{y}}} > 1/2 + \negl[\lambda]]
\end{equation*}
By a union bound, a bound on the desired probability is given by
\begin{equation*}
    \label{eq:initialprob}
    \Pr[G \text{ is correct and } p_{T,\widetilde{\boldsymbol{y}}} > 1/2 + \negl[\lambda]] > p_G + p_{\mathrm{good}} - 1
\end{equation*}
Afterward, we try to write $p_{\mathrm{good}}$ in terms of $p_T$. Let $S$ be the set of $\widetilde{\boldsymbol{y}}$ values for which $p_{T,\widetilde{\boldsymbol{y}}} > 1/2 + \negl[\lambda]$. By definition, it is known that with probability $p_{\mathrm{good}}$, the prover samples a $\widetilde{\boldsymbol{y}} \in S$ so that they pass the verifier's equation test with probability at least $1/2 + \negl[\lambda]$ and at most $1$. In addition, note that with probability $1 - p_{\mathrm{good}}$, the prover samples a $\widetilde{\boldsymbol{y}} \notin S$ so that they pass the verifier's equation test with probability at most $1/2$. Therefore, we can see that the probability that the \(\mathsf {PPT}\) prover passes the equation test round is at most the convex mixture of these two cases, i.e.
\begin{equation*}
    p_T < 1 \cdot p_{\mathrm{good}} + 1/2 \cdot (1 - p_{\mathrm{good}})
\end{equation*}
Solving for $p_{\mathrm{good}}$, we can obtain
\begin{equation*}
    p_{\mathrm{good}} > 2p_T - 1
\end{equation*}
By substituting this equation, we then get
\begin{equation*}
    \label{eq:breakahb}
    \Pr[G \text{ is correct and } p_{T,\widetilde{\boldsymbol{y}}} > 1/2 + \negl[\lambda]] > p_G + 2p_T - 2
\end{equation*}
It is known that this probability on the left hand side is indeed the probability of breaking the adaptive hardcore bit (AHB) property, which we know \cite{BCMVV18} must satisfy
\begin{equation*}
    \Pr[G \text{ is correct and } p_{T,\widetilde{\boldsymbol{y}}} > 1/2 + \negl[\lambda]] < \negl[\lambda]
\end{equation*}
Hence, we can obtain the desired inequality
\begin{equation*}
    p_G + 2p_T - 2 < \negl[\lambda]
\end{equation*}
  
\end{proof}

\section{ Conclusion and open problems} \label{Dis}

In this paper, we consider the 2-to-1 NTCF$^1_2$ extension and construct $\kappa$-to-1 NTCF$^1_{\kappa}$ with $\kappa = \poly$, assuming the QLWE assumption. Our solution manifests an interesting connection between the NTCF$^1_{\kappa}$ and the extrapolated dihedral coset state. Based on this, we demonstrate that the NTCF$^1_2$ (resp. NTCF$^1_{\kappa}$) based on QLWE is an explicit description of the black-box function of the DCP (resp. EDCP).

Thus, we firstly instantiate a $\ZZ^n_q$-DCP (resp. EDCP) oracle by using NTCF$^1_2$ (resp. NTCF$^1_{\kappa}$), thereby directly deriving quantum reductions from the LWE to the $\ZZ^n_q$-DCP and the EDCP. Remarkably, our solution presents a simpler mathematical abstraction of the reduction from LWE to DCP and can be visualized as an improved Balls' intersection method compared to \cite{BKSW18}, i.e., replacing the uniformly distributed error ball with a Gaussian distributed error ball.

Moreover, we propose a two-round protocol for proving the quantumness based on NTCF$^1_{\kappa}$. In order to run the equation test on a superposition over $\kappa$ preimages produced by NTCF$^1_{\kappa}$, we further design an additional quantum polynomial-time reducing procedure. Due to the application of NTCF$^1_{\kappa}$ instead of the NTCF$^1_{\kappa}$, our protocol could allow the prover to produce a superposition over $\kappa$ preimages under the ideal case.

Next, let us focus on the efficient generation of cryptographically certified random bits via the NTCF$^1_{\kappa}$. As we know, since the \(\mathsf {QPT}\) prover can produce a superposition over $\kappa$ preimages via the NTCF$^1_{\kappa}$, a natural idea is to design a certifiable quantum randomness generation protocol for generating approximately $\bigTheta{\log n}$ bits of randomness. Unfortunately, our NTCF$^1_{\kappa}$-based interactive protocol only serves as a proof of quantumness, since passing the equation test by using the post-processing reducing procedure, our protocol only implies that the prover actually holds the superposition over two preimages, i.e. $\frac{1}{\sqrt{2}}(\ket{0}\ket{\bar{\boldsymbol{x}}_0}+\ket{1}\ket{\bar{\boldsymbol{x}}_1})$ rather than $\frac{1}{\sqrt{\kappa}}\sum_{b} \ket{b}\ket{\boldsymbol{x}_b}$ with $b\in \{0,...,\kappa-1\}$. Thus, the final protocol only provides 1 bit of certifiable randomness. Therefore, it is unclear how to directly test the quantum prover’s behavior (honest or not) for the case when $\kappa$-to-1 NTCFs are used. Based on the above, we give an interesting open question 
on how to use the NTCF$^1_{\kappa}$ to further achieve a classically certifiable quantum randomness generation protocol, which would be a very important application of the many-to-one trapdoor claw-free functions in the future.

Finally, we suggest another important question in the context of adopting NTCF$^1_{\kappa}$ to prove quantumness. Similar to the NTCF$^1_{2}$ introduced in \cite{BCMVV18}, our construction of NTCF$^1_{\kappa}$ also relies on the super-polynomial LWE modulus. Since most application of LWE has the polynomial modulus, the next question is whether it is possible to reduce the super-polynomial LWE modulus, thereby allowing us to construct NTCFs based on the hardness of LWE problem with polynomial modulus.


\section*{Declaration of competing interest}
The authors declare that they have no known competing financial interests or personal relationships that could have appeared to influence the work reported in this paper.

\section*{Data availability}
Data sharing is not applicable to this article as no datasets were generated or analyzed during the current study.

\section*{Acknowledgments}
The authors deeply thank Weiqiang Wen for many insightful exchanges and discussions.
This work was supported by the National Key Research and Development Program of China (No. 2022YFB2702701), and the National Natural Science Foundation of China (NSFC) (No. 61972050).



\end{document}